\documentclass[%
reprint,
amsmath,amssymb,
aps,
prb,
]{revtex4-2}
\usepackage{graphicx}
\usepackage{dcolumn}
\usepackage{bm}
\usepackage{xcolor}
\usepackage{comment} 
\usepackage{placeins}
\newcommand{\Tbkt}{T_\mathrm{BKT}}
\newcommand{\Tczero}{T_{c0}}

\newcommand{\RN}{R_\mathrm{N}}
\newcommand{\Rsq}{R_\square}
\newcommand{\Rc}{R_\mathrm{c}}

\newcommand{\red}[1]{{\color{red}#1}}

\newcommand{\beginsupplement}{%
	\setcounter{table}{0}
	\renewcommand{\thetable}{S\arabic{table}}%
	\setcounter{figure}{0}
	\renewcommand{\thefigure}{S\arabic{figure}}%
	\renewcommand{\theequation}{S.\arabic{equation}}
}

\newcommand{\beginextendeddata}{%
	\setcounter{figure}{0}
	\renewcommand{\thefigure}{\arabic{figure}}
	\renewcommand{\figurename}{Extended Data Fig.}
	\setcounter{table}{0}
	\renewcommand{\thetable}{\arabic{table}}
	\renewcommand{\tablename}{Extended Data Table}
}

\newcommand{\stopextendeddata}{%
	\setcounter{figure}{0}
	\renewcommand{\thefigure}{\arabic{figure}}
	\renewcommand{\figurename}{Fig.}
	\setcounter{table}{0}
	\renewcommand{\thetable}{\arabic{table}}
	\renewcommand{\tablename}{Table}
}

\newcommand{\br}{{\boldsymbol{r}}}
\newcommand{\Js}{J_\text{s}}

\begin{document}
	
	\preprint{APS}
	\title{Origin of the superconductor-insulator transition in disordered two-dimensional films}
	
	\author{Alexander Weitzel,$^1$ Lea Pfaffinger,$^1$ Animesh Panda,$^2$ Arpan Das,$^2$ Ilaria Maccari,$^3$ Simon Reinhardt,$^{1*}$  Sven Linzen,$^4$ Evgenii Il'ichev,$^4$ Nicola Paradiso,$^{1,5}$ Ferdinand Evers,$^{2,5}$  Christoph Strunk$^{1,5}$}
	\affiliation{$^1$Institute for Experimental and Applied Physics, University of Regensburg, 93040 Regensburg, Germany}
	\affiliation{$^2$Institute for Theoretical Physics, University of Regensburg, 93040 Regensburg, Germany}
	\affiliation{$^3$Institute for Theoretical Physics, ETH Zurich, 8093 Zurich, Switzerland}
	\altaffiliation{Present address: School of Applied and Engineering Physics, Cornell University, Ithaca, NY, 14853, USA}
	\affiliation{$^4$Leibniz Institute of Photonic Technology, 07745 Jena, Germany}
	\affiliation{$^5$Halle-Berlin-Regensburg Cluster of Excellence CCE, University of Regensburg, 93040 Regensburg, Germany}
	
	\date{\today}

	\begin{abstract}
		Theory predicts the superconductor-to-insulator transition (SIT) to emerge from the competition between Anderson localization, which tends to localize single-particle wavefunctions, and superconductivity,
		which establishes long-range correlations in the superconducting order parameter. 
		In two-dimensional (2D) superconducting films, the transition temperature $T_\text{c}$ at which resistance vanishes, $\Rsq(T_\text{BKT}){=}0$, is set by the Berezinskii-Kosterlitz-Thouless (BKT) mechanism and satisfies $T_\text{BKT}< \Tczero$, where $\Tczero$ is the mean-field transition temperature. In weakly disordered samples $T_\text{BKT}\lesssim \Tczero$, whereas increasing disorder 
		drives $T_\text{BKT}\ll \Tczero$ near the SIT. Whether the finite-temperature transition retains its BKT character throughout this crossover remains an open question. 
		Here, we investigate the evolution of both  sheet resistance $R_\Box(T)$ and superfluid stiffness $\Js(T)$ over a wide range of disorder strength $W$. 
		We establish that even near the SIT, the finite-temperature transition from the superconducting to the resistive state remains of BKT type. 
		However, as disorder approaches the critical value, the zero temperature superfluid phase stiffness, $\Js(0)$, is found to vanish rapidly
		while $\Tczero$ remains finite, which we attribute to quantum phase fluctuations as the drive for the zero-temperature transition.
		Three decades after its experimental discovery by Haviland, Liu, and Goldman, our measurements clarify the origin of the SIT in 2D films.
		
	\end{abstract}

	\maketitle
	
	
	In strongly disordered superconducting films, the  localization of Cooper pairs leads to the superconductor-insulator transition (SIT), which is considered a prime example of a quantum phase transition 
	that occurs at zero temperature at a critical strength of disorder potential $W$  
	\cite{Sondhi1997-hr,Gantmakher2010-zs}. 
	The nature of the quantum fluctuations driving this transition is still not fully understood. Disorder can suppress superconductivity through two distinct mechanisms that act, respectively, on the modulus and phase of the complex superconducting order parameter $\Delta(r) = |\Delta(r)|e^{i\varphi(r)}$. The first scenario   (“fermionic”) emphasizes that the dynamic screening of electrons is slowed down as the disorder becomes stronger, implying weaker pairing and, eventually, a global breakdown of $|\Delta(\br)|$ \cite{Finkelstein1994-ub}. 
	In the second scenario (“bosonic”), $|\Delta(\br)|$ is reduced as well, but remains non-vanishing on average also on the insulating side of the SIT. Here, superconductivity breaks down because with increasing disorder the global phase-stiffness $\Js$ reduces, enabling quantum fluctuations of the superconducting phase $\varphi(r)$ to proliferate and destroy the long-range order in $\Delta(\br)$ \cite{Fisher1989-ib,Fisher1990-cm,Fisher1990-om,Emery1995-yi}. 
	This scenario should be distinguished from the BKT-mechanism \cite{Berezinskii1972, Kosterlitz1973-zt,Kosterlitz1974-mc}. It describes how superconductivity vanishes in a thin film due to increased thermal phase fluctuations; for this case, a quantitative theory has been  devised \cite{HalperinNelson_1979,AHNS_1980}.

	Despite more than three decades of intensive research, two most pressing questions in this field are still open: i) is the SIT  of bosonic or fermionic type, and ii) what is the fate of the BKT transition at the SIT?

	\begin{figure}[t]
		\includegraphics[width=0.48\textwidth]{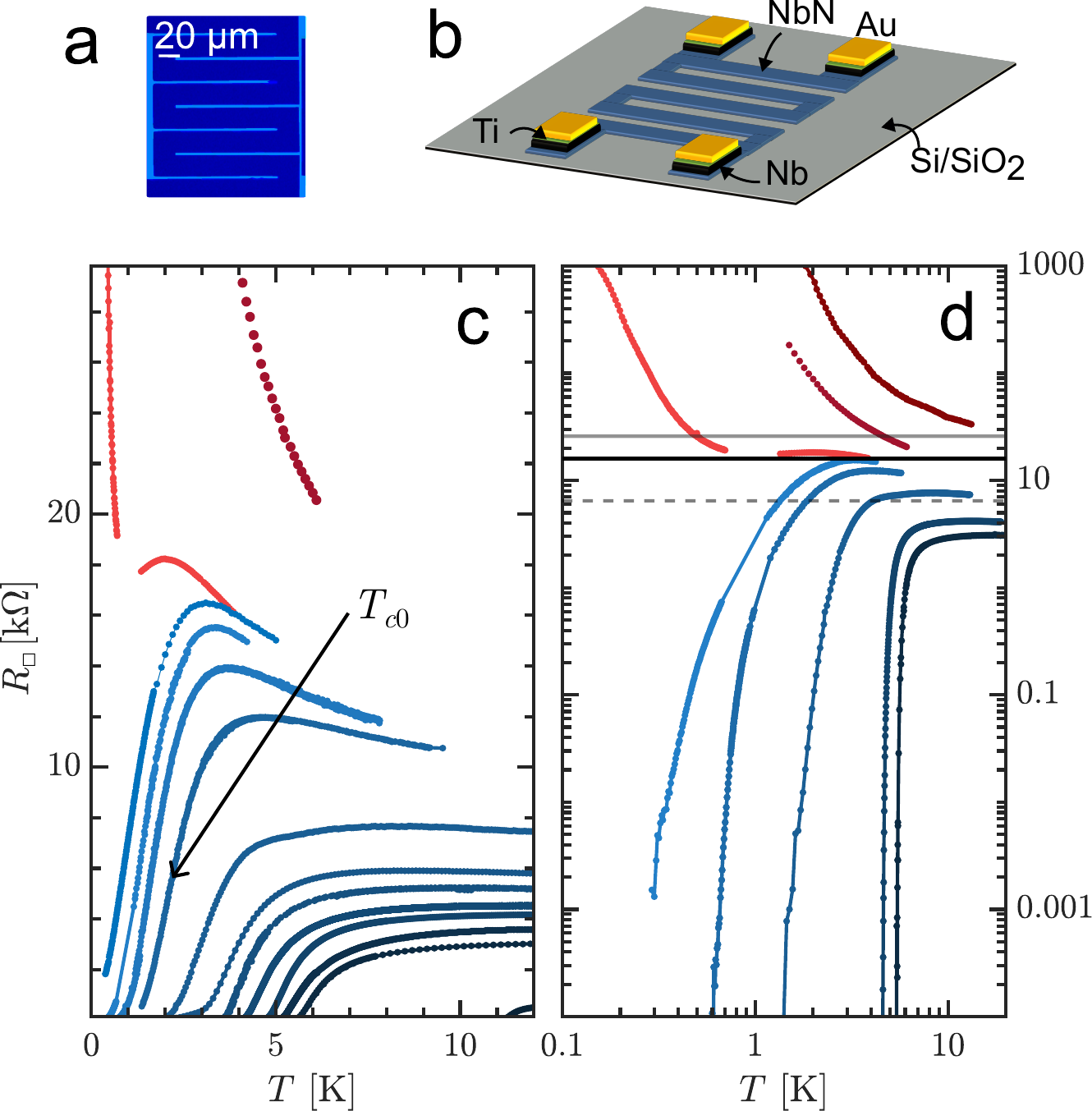}
		\caption{{\bf Setup and DC resistance: a)} Optical micrograph of a typical device: argon ion milling is used to etch the 3 ~nm thick NbN film into a meander shape. {\bf b)} Schematic of the 4-terminal contact design. {\bf c)} Sheet resistance vs.~temperature of 13 NbN-film with room-temperature resistance up to 3.5\,k$\Omega$. The normal state resistance $\RN$ including all quantum corrections is assigned to the maximum of $R_\Box(T)$. {\bf d)} Selected $R_\Box(T)$-curves on a logarithmic scale. 
		}
		\label{fig:}
	\end{figure}

	In this work, we present a systematic study of the evolution of the  BKT transition at nearly critical disorder $W_\text{c}$. We measure $\Rsq(T)$ and $\Js(T)$ all the way up to $W_\text{c}$  and explore the complete evolution of  $\Tczero$, $\Tbkt$ and $\Js(0)$ with disorder.  
	Starting from lower disorder, the initial suppression of $\Js(0)$ and $\Tbkt$ with increasing $W$ follows 
	that of $\Tczero$, as one would expect within the fermionic mechanism described by Finkel'stein~\cite{Finkelstein1994-ub}. Very close to the SIT, however, we observe a sharp drop in $\Js(0)$ and $\Tbkt$,  below the expectation from mean-field theory, while $\Tczero$ continues to follow  the Finkel'stein theory. Our work allows for the so far elusive discrimination of the fermionic and bosonic mechanisms leading to the SIT and provide unambiguous evidence for quantum phase fluctuations close to critical disorder.  
	%

	\noindent\textbf{Resistive transition near the SIT}\\
	An optical micrograph and a schematic of the sample design are shown in Figs.~\ref{fig:}a,b. In total, we measured fifteen  superconducting and three  insulating samples. We present DC-resistance vs.~temperature, $R_\Box(T)$,  in Fig.~\ref{fig:}c. Figure~\ref{fig:}d shows the same $\Rsq(T)$ traces on a logarithmic scale. We observe a direct SIT without intermediate metallic phase. Higher sheet resistance in the normal state corresponds to a lower superconducting transition temperature, as is typical for homogeneously disordered films \cite{Finkelstein1994-ub,Shahar1992,Sacepe2008,Marrache-Kikuchi2008-hh,Linzen_2017}. Around $T_\mathrm{max}\simeq2.5\, \Tczero$, the superconducting samples display a resistance maximum (not visible in Fig.~\ref{fig:}c,d 
	for the lower resistance films), which arises from the competition of superconducting fluctuations with the normal-state quantum corrections to the resistance \cite{Baturina_2012}. Between $T_\mathrm{max}$ and 300\,K the resistance decreases with increasing $T$ logarithmically by a factor three or more, and is dominated by the disorder-enhanced electron-electron interaction \cite{Altshuler1982-hs}. 
	
	Following the work of König {\it et al.} \cite{Koenig2015}, which takes into account the effect of all quantum corrections on $\Rsq(T)$ as well as quantum fluctuations on the BKT-transition, the level of disorder is defined via $\RN=\Rsq(T_\mathrm{max})$. A mapping between $\RN$ and $W$ is described in \cite{Supplement}. 
	The SIT occurs at a critical value of $\RN$: 16\,k$\Omega<\Rc<18$\,k$\Omega$ (black line in Fig.~\ref{fig:}d).   Below $T_\mathrm{max}$ and above $\RN/2$, the strongly broadened $\Rsq(T)$ curves are dominated by amplitude fluctuations of the SC order parameter close to $\Tczero$ \cite{Weitzel2023}. We assign the inflection point of the $\Rsq(T)$-curve to the  pairing temperature $\Tczero$ \cite{Supplement}. Very similar values of $\Tczero$ are obtained from a fit of $\Rsq(T)$ to the theory of amplitude fluctuations  \cite{Weitzel2023}.  
	While a superfluid condensate is already formed  at $\Tczero$, $\Rsq(T)$ does not vanish above $\Tbkt$: resistance there is described by the Halperin-Nelson formula  \cite{HalperinNelson_1979}:
	\begin{equation}\label{eq:SRC}
		\Rsq(T)=C\;\exp\big(-b/\sqrt{T/\Tbkt-1}\big)\,.
	\end{equation}
	Fitting this expression to the $\Rsq(T)$-data provides experimental access to  $\Tbkt$, below which the zero-resistance state is established \cite{Baturina_2012}. We find that Eq.~\ref{eq:SRC} results in an excellent fit even very close to the SIT (Extended Data Fig.~1 and \cite{Supplement}), with good agreement between data and theory over $3-4$ orders of magnitude in $\Rsq$. 
	For $\RN\gtrsim16\,\text{k}\Omega$ the $T$-dependence turns insulating while no intervening metallic phase \cite{Kapitulnik2019-te} is observed.\\[1mm]

	\begin{figure*}[t]
		\includegraphics[width=\textwidth]{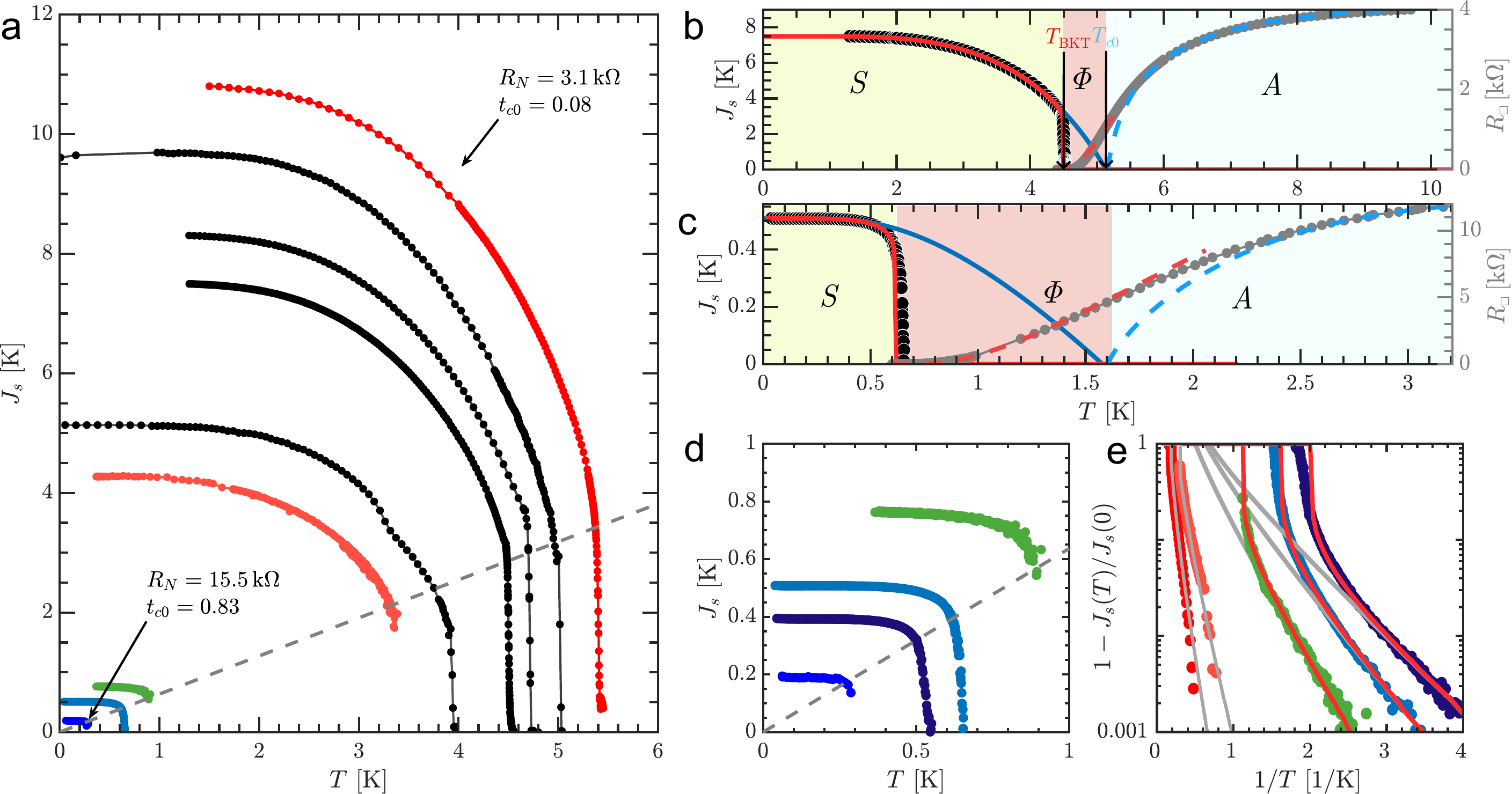}
		\caption{{\bf Superfluid phase stiffness: a)} Measured $\Js(T)$ for selected films. The dashed line indicates the universal transition point at $\Js(\Tbkt)=2\Tbkt/\pi$. {\bf b)} Stiffness and resistance for $\RN=4.1\,\text{k}\Omega$. Solid lines are fits to $\Js(T)$ according to the BCS (blue) and BKT (red) theories. Dashed lines are fits describing phase (red, Eq.~\ref{eq:SRC}) and amplitude fluctuations. Colored shades denote areas that are fully superconducting (S) or dominated by  phase ($\Phi$) and amplitude (A) fluctuations, respectively. {\bf c)} Same for $\RN=12.3\,\text{k}\Omega$, close to the SIT. {\bf d)} Zoom of low-temperature range in {\bf a)}. Colored curves correspond to those in {\bf a}.  {\bf e)} Exemplary Arrhenius plots of the difference $1-\Js(T)/\Js(0)$ for $\RN=3.1,5.2,12.0,12.3,13.0$ k$\Omega$ (left to right).  Grey lines are fits to Eq.~\ref{eq:J_BCS}, while red lines corresponds to the RG-results. These fits provide experimental values for the characteristics energy scales $E^\ast\text{ and }\mu(0)$ (see text).
		}
		\label{fig:Js(T)}
	\end{figure*}

	\noindent\textbf{Superfluid phase stiffness near the SIT}\\
	Next, we investigate the signatures of the BKT-transition in the superfluid phase stiffness $\Js(T)$ which is universally related to the sheet kinetic inductance $L_{\square}$ via
	\begin{align}\label{eq:Js}
		\Js(T)\ =\ \frac{\hbar^2d}{4e^2k_\mathrm{B}\mu_0\lambda^2(T)}\ =\ \frac{\hbar^2}{4e^2k_\mathrm{B} L_{\square}(T)}\;, 
	\end{align}
	where  $h$ is Planck's constant, $e$ the electron charge, $\lambda$ the magnetic penetration depth and $k_\mathrm{B}$ being Boltzmann's constant.  
	According to BCS-theory, the  temperature dependence of $\Js(T)$ is given by
	\begin{align}\label{eq:J_BCS}
		\Js(T) =%
		\Js(0)\frac{\Delta(T)}{\Delta(0)}\cdot\tanh\left[\frac{E_\mathrm{g}(T)}{2k_\mathrm{B} T}\right]\;.
	\end{align}
	where $E_\mathrm{g}(T)$ is the quasiparticle gap and 
	\begin{equation}\label{eq:J_s0}
		J_\mathrm{BCS}(0)=\frac{\pi1.76\,\hbar\,\Tczero}{4e^2 \RN}\;.
	\end{equation} 
	The form of Eq.~\ref{eq:J_BCS} is chosen such that $E_\mathrm{g}(0)$, $1.76\,k_\mathrm{B} \Tczero$ and $\Delta(0)$ can differ from each other  as expected for $\RN\gtrsim h/4e^2$ \cite{Ghosal1998,Ghosal2001,Stosiek2020, Weitzel2023}.
	The BCS stiffness in Eq.~\ref{eq:J_BCS} is renormalized by the presence of vortex-antivortex (VAV) pairs and discontinuously drops to zero at the universal BKT line $\Js(T_\mathrm{BKT})=2\,T_\mathrm{BKT}/\pi$ where VAV pairs dissociate. 
	
	Figure~\ref{fig:Js(T)}a  shows $\Js(T)$ for a representative selection of the fifteen superconducting samples on a linear scale. All the way down to the SIT, all curves initially follow  Eq.~\ref{eq:J_BCS} (solid grey lines) and drop close to the universal line (dashed grey lines in Fig.~\ref{fig:Js(T)}a,d).  
	Note that $\Js(0)$ varies with $\RN$ over nearly two orders of magnitude. For the highest $\RN\simeq16~\text{k}\Omega$ the width of the thermal phase fluctuation regime amounts to an unprecedented value of 80\% of $\Tczero$.
	
	The agreement of our data with the BCS and BKT theories is exemplarily shown in Figs.~\ref{fig:Js(T)}b,c. At lower $T$, $\Js(T)$ follows Eq.~\ref{eq:J_BCS}. The width of the thermal phase fluctuation regime (shaded in rose) is quantified by the reduced BKT temperature $t_{c0}=(\Tczero-\Tbkt)/\Tczero$;  for $\RN\simeq4\,\text{k}\Omega$ we obtain $t_{c0}  \simeq {0.13}$ (pink area in Fig.~\ref{fig:Js(T)}b,c). The extrapolated BCS fit of $\Js(T)$ (dark blue line) and the fit to amplitude fluctuations (light blue dashed line) 
	vanish at very similar values of $\Tczero$. 
	Similarly, the measured $\Js(T)$ (black dots) and $\Rsq(T)$-curves vanish at $\Tbkt$, together with the fit to Eq.~\ref{eq:SRC} (red dashed line) and a numerical renormalization group (NRG) calculation (solid blue line) \cite{Supplement}. 
	
	Closer to the SIT, i.e., at  $\RN\simeq12~\text{k}\Omega$ (Fig.~\ref{fig:Js(T)}c), the width of the phase fluctuation regime is much wider {($t_{c0}\simeq0.65)$}. On a linear scale and starting from low $T$, both $\Js(T)$ and $\Js^\text{BCS}(T)$ appear essentially constant. The only visible $T$-dependence results from the renormalization of $\Js(T)$ near the universal line.  Near $\Tczero$, $\Rsq(T)$ is  linear, similar to the observations in InO$_x$ \cite{Charpentier2025}. Also in this case, the value of $\Tczero$ extracted from 
	$\Rsq(T)$ reasonably agrees with the inflection point of $\Rsq(T)$, where the phase  (positive curvature) and amplitude  (negative curvature) fluctuation regimes meet (Fig.~\ref{fig:sup:T_c0_analysis}).
	Consistent values of $\Js(T)$ can also be extracted from non-linear $IV$-curves (\cite{Weitzel2023}, Extended Data Fig.~\ref{fig:ext_data:non_linear_transport}).
	
	
	Besides $\Js(0)$ and $\Tbkt$, we can extract two more energy scales from $\Js(T)$: an activation energy, $E^\ast$, which determines the exponentially small value of $1-\Js(T)/\Js(0)$ at low $T$, and the chemical potential of VAV-pairs $\mu$. 
	In Fig.~\ref{fig:Js(T)}e we plot $1-\Js(T)/\Js(0)$ for five samples with different $\RN$ in an Arrhenius fashion, i.e., vs.~$1/T$. Using $E^\ast$ and $\Js(0)$, we fit the data in Fig.~\ref{fig:Js(T)}e to {\bf (i)} Eq.~\ref{eq:J_BCS} (grey lines). Here, we use Eq.~\ref{eq:J_BCS} in a  heuristic way, without assuming $E^\ast=E_\mathrm{g}$. and {\bf (ii)} using the NRG-procedure \cite{Benfatto2009,Maccari2017} with Eq.~\ref{eq:J_BCS} as starting point (red lines). The initial values of $\mu(T=0,\RN)$ entering the NRG are determined by adjusting the position of the universal jump in $\Js(T)$ calculated using NRG to the data.
	
	\begin{figure*}[htbp]
		\includegraphics[width=13.5cm]{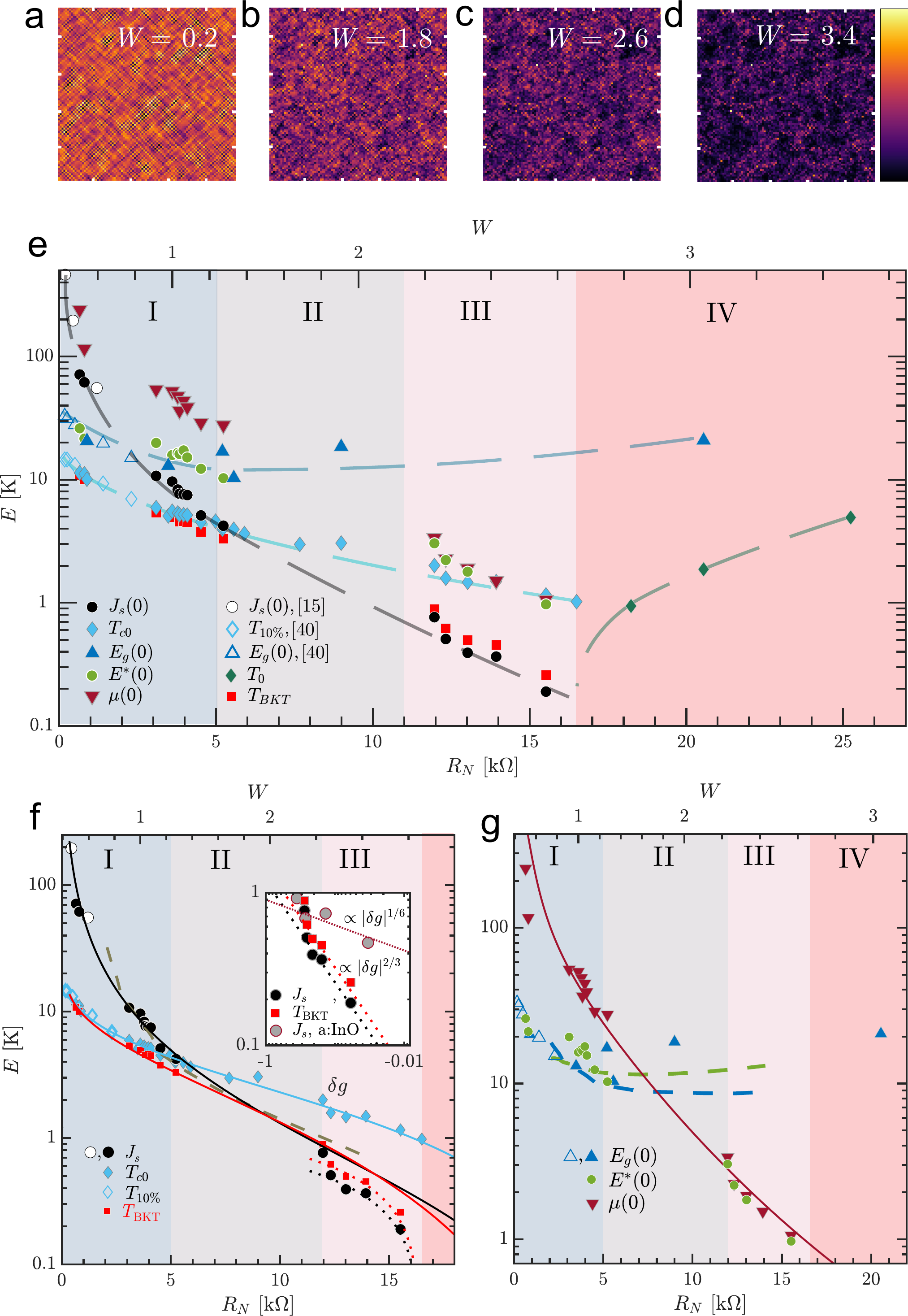}
		\caption{{\bf Evolution of energy scales with disorder: a-d)}  Simulated maps of the pairing potential $\Delta(0)$ based on a fully self-consistent mean-field calculation on a lattice with 96x96 sites corresponding to regimes I-IV in (e), respectively (Supplement, Sec.~\ref{sec:sup:modeling}). The color bar corresponds to $\Delta/t$= [0.1375 0.1825], [0.025 0.25], [0.01 0.3], [0 0.375] for a,b,c,d, respectively.
			{\bf e)} Relevant energy scales in the problem (see text). Long dashed lines are guides to the eyes.
			{\bf f)} Comparison of experiment and theory for $\Tczero$ and Finkel'stein (Eq.~\ref{eq:finkelstein}, light blue line), for $\Js(0)$ based on Mattis-Bardeen (Eq.~\ref{eq:J_BCS}, black line), and  for $\Tbkt$ based (Eq.~\ref{eq:Tbkt}, light red line).	Inset: Double-logarithmic plot for $\Js(0)$ and $\Tbkt$ vs.~reduced conductance $\delta g=(R_c-\RN)/R_c^2$. The red and black dotted lines denote power-law near the critical resistance $R_c$ with exponent $0.67$ while the vertical shift corresponds to the factor $2/\pi$ in the universal BKT-relation. The grey circles indicate $\Js(0)$ for a-InO$_x$ \cite{Charpentier2025}. 
			{\bf g)} Characteristic energies $E^\ast$, $\mu$, and $E_g$ vs.~$\RN$. The dashed lines denote $E^\ast$ (green) and $E_g$ (blue) from fully self-consistent simulations of the system \cite{Supplement}. The  vortex chemical potential $\mu(\RN)$ extracted from the NRG-fits (dark red triangles) agrees well with ${\mu}=\pi^2/2\cdot \Js(0)$ from the combination of the $xy$-model with an interpolation of the measured values of $\Js(0)$ (dark red line). 
		}
		\label{fig:analysis}
	\end{figure*}

	\noindent\textbf{Energy scales near the BKT transition}\\
	The central result of our work is a systematic comparison of the relevant energy scales over a wide range of disorder.  Four regimes  can be distinguished: (I) the BCS-like suppression of stiffness and fermionic suppression of  $|\Delta(T=0)|$ with increasing disorder, (II) a crossover from quasiparticles to a different type of excitation, (III) the regime of strong quantum phase fluctuations and, finally, (IV) the insulating phase. 
	The color plots in Fig.~\ref{fig:analysis}a-d map out numerical simulations of the local distribution of $|\Delta(0)|$ \cite{Stosiek2020,Supplement}. An overall decrease of $\langle|\Delta(0)|\rangle$ is seen together with a progressively more pronounced emergent granularity of the pair potential \cite{Ghosal1998,Ghosal2001} that leads to a filamentary structure. In region~II, a crossover of the $\Delta$-distribution from a peak centered around a finite value of $\langle|\Delta(0)|\rangle$ to a  peak at $\langle|\Delta(0)|\rangle=0$ is predicted \cite{Dieplinger}. 	In Figure~\ref{fig:analysis}e, we give an overview of all relevant energy scales (in units of temperature) in our strongly disordered NbN films.  To highlight two core results of our work, we extract  subsets of the data in Fig.~\ref{fig:analysis}e in Figs.~\ref{fig:analysis}f and~\ref{fig:analysis}g. 
	
	\noindent\textbf{Comparison to theory:}  \\ Figure~\ref{fig:analysis}f displays the relation between the pairing temperature $\Tczero$ (light blue diamonds), the stiffness $\Js(0)$ (black circles) and the BKT temperature $\Tbkt$ (red squares) in both  theory and experiment. In regime~I, $\Tczero$ and $\Tbkt$ are very close, because $\Js(0)$ significantly exceeds $\Tczero$. In regime~II, $\Tczero$ and $\Tbkt$ separate from each other, while in regime~III,  $\Tbkt$ and $\Js(0)$ are close and lie significantly below $\Tczero$. The position of the crossover coincides with the minimum of the quasiparticle excitation energy $E_\mathrm{g}$ (green circles in Fig.~\ref{fig:analysis}g, obtained from tunnel spectroscopy \cite{WeitzelIII}) and the crossover of the $\langle|\Delta(0)|\rangle$-distribution in Fig.~\ref{fig:analysis}b \cite{Dieplinger}. 
	The blue line represents a fit of the Finkel'stein theory, which accounts for the fermionic suppression of $\Tczero$ by strong disorder (Extended Data Fig.~\ref{fig:ext_data:finkelstein} and \cite{Supplement}). Inserting the Finkel'stein fit into Eq.~\ref{eq:J_s0}, we can derive the BCS expectation for $\Js(0)$ (black line in Fig.~\ref{fig:analysis}f).
	Finally, inserting the Finkel'stein fit for $\Tczero(\RN)$ into the relation \cite{Larkin2005,Koenig2015} 
	\begin{equation}\label{eq:Tbkt}
		T_\mathrm{BKT}(\RN)=T_\mathrm{c0}(\RN)\big(1-4\,Gi(\RN)\big)\;
	\end{equation}  
	provides an expectation for $\Tbkt(\RN)$ (red line in Fig.~\ref{fig:analysis}f).
	Here $Gi(\RN)=e^27\zeta(3)\RN/(\pi^3h)\simeq 0.27\,\RN e^2/h$ is the Ginzburg-Levanyuk number, which quantifies the regime of Gaussian fluctuations. 
	In addition, the dashed grey line depicts $J_s(R_N)$ obtained from numerical simulations that underlie  Figs.~\ref{fig:analysis}a-d. The functional form of the numerical results is close to that of Eq.~\ref{eq:J_BCS}. 
	
	We find good agreement between these theories (solid lines) and experiment in regimes I and II.  It is seen that $\Js(0)$ (black circles in Fig.~\ref{fig:analysis}f) initially follows the BCS-prediction (black line) $J^\text{BCS}_s(0)$ as obtained from Eq.~\ref{eq:J_s0} in regimes~I and~II.  
	In regime~III, i.e.~beyond $\RN\simeq12\,\text{k}\Omega$,  $\Js(0)$ falls significantly below its mean-field value (Eq.~\ref{eq:J_s0}).  We attribute this suppression to quantum phase fluctuations of the order parameter, possibly resulting from emergent granularity of $|\Delta|$. 
	Hence, we can separate for the first time fermionic and bosonic effects in the suppression of phase stiffness \cite{Emery1995-yi}. 
	%
	Similarly,  Eq.~\ref{eq:Tbkt} describes the measured variation (dots) of $\Tbkt$ rather well at lower values of $\RN$. 
	In region~III, also $\Tbkt$ drops faster than expected, while maintaining the universal relation $\Tbkt=2\Js(\Tbkt)/\pi$.
	For the quantum phase fluctuation regime, it was predicted that  $\Js(0)\propto (\delta g)^\nu$ and $\Tbkt\propto (\delta g)^{z\nu}$ where $\delta g=(\RN-R_c)/R_c^2$;   $\nu$ being the standard critical exponent  for the correlation length and $z$ the dynamical critical exponent \cite{Fisher1989-ib,Fisher1990-cm,Fisher1990-om}.
	Close to the SIT, we find approximately  $\Js(0)\simeq \Js(\Tbkt)=\pi\Tbkt/2$ implying $z\simeq1$. Black and red dotted lines indicate that our data are consistent with the value $\nu=0.67$. Investigating superconducting oxide interfaces \cite{Caviglia2008-jp}, an exponent $z\nu=0.67$ was obtained for $\Tbkt\propto (\delta g)^{z\nu}$ (see also \cite{Parendo2005-vq,Aubin2006-bg}).  
	
	\textbf{Crossover from charge to collective excitations?} \\
	In Figure~\ref{fig:analysis}g, we plot the excitation energies $E^\ast$ extracted from the Arrhenius plots (Fig.~\ref{fig:Js(T)}e). Obviously, Bogoliubov quasiparticles result from thermal pair breaking across the spectral gap $E_g(\RN)$ (dark blue triangles). Using scanning tunneling spectroscopy (STS),  $E_g$ was measured near the SIT for NbN \cite{Chand2012-kn,Mondal_2011a, Noat2013, Carbillet2016}. To measure $E_g$ of our films, we fabricated planar tunnel junctions for tunneling spectroscopy \cite{WeitzelIII}.
	In agreement with the STS data, we find that after an initial decrease in regime~I, the resulting $E_g(\RN)$ features a shallow minimum in regime~II and persists in the insulating regime~(IV). This behavior is similar to the observation for InO$_x$ \cite{Charpentier2025}. 
	In addition, in Figs.~\ref{fig:analysis}e,g we display the VAV chemical potential $\mu$, also referred to as the vortex core energy $\mu(\RN)$. In the low-$T$ limit his is the energy needed to excite a bound VAV pair at  minimal distance. 
	As indicated by the red line, the measured $\mu(\RN)$ agrees well with simulations of the $xy$-model \cite{Benfatto_2013}, where the ratio $\tilde{\mu}=\mu(T)/\Js(T)=\pi^2/2\simeq4.96$ is independent of both $T$ and $\RN$. In contrast to our 2D films, power-law behavior of $\Js(T)$ in InO microstrip resonators. Nevertheless, a characteristic temperature $T_0\simeq 5.5\Js$ was reported at low $T$ \cite{Khvalyuk2024-mr}. This value is very close to our above result for $\mu(0)$; thus we may speculate that both scales are related. 	
	In regime~I, we observe 
	$E^\ast\simeq E_g<\mu$ implying that the Bogoliubov quasiparticles dominate the thermal depletion of the superfluid stiffness. In regime~II, $E^\ast$ breaks away from $E_g$. Very close to the SIT we find $E^\ast\simeq\mu<E_g$. In the numerical simulations (dashed lines in Fig.~\ref{fig:analysis}g), in contrast, $E^\ast$ remains close to $E_g$ over the whole disorder range, which is expected, because vortices are not included in these simulations.
	The observation that the activated behavior of $\Js(T)$ crosses over from that of quasiparticles to another type of excitation has implications for the design of quantum circuits based on disordered superconductors.
	
	The quantitative agreement between $E^\ast$ and $\mu$ in region~III is striking. At first sight it is tempting to associate $E^\ast$ with VAV-pairs and simply view $\mu$ as their excitation energy. However, the situation is even more interesting than that: formally, $\mu$ enters our  analysis as a starting condition of a renormalization group flow. The generic flow pattern does not foresee $\mu$ resembling an energy scale like $E^\ast$ that only emerges at low temperature. Therefore, their apparent quantitative agreement in region~III remains an intriguing puzzle. 
	\\[1mm]
	
\begin{figure}[t]
		\includegraphics[width=0.4\textwidth]{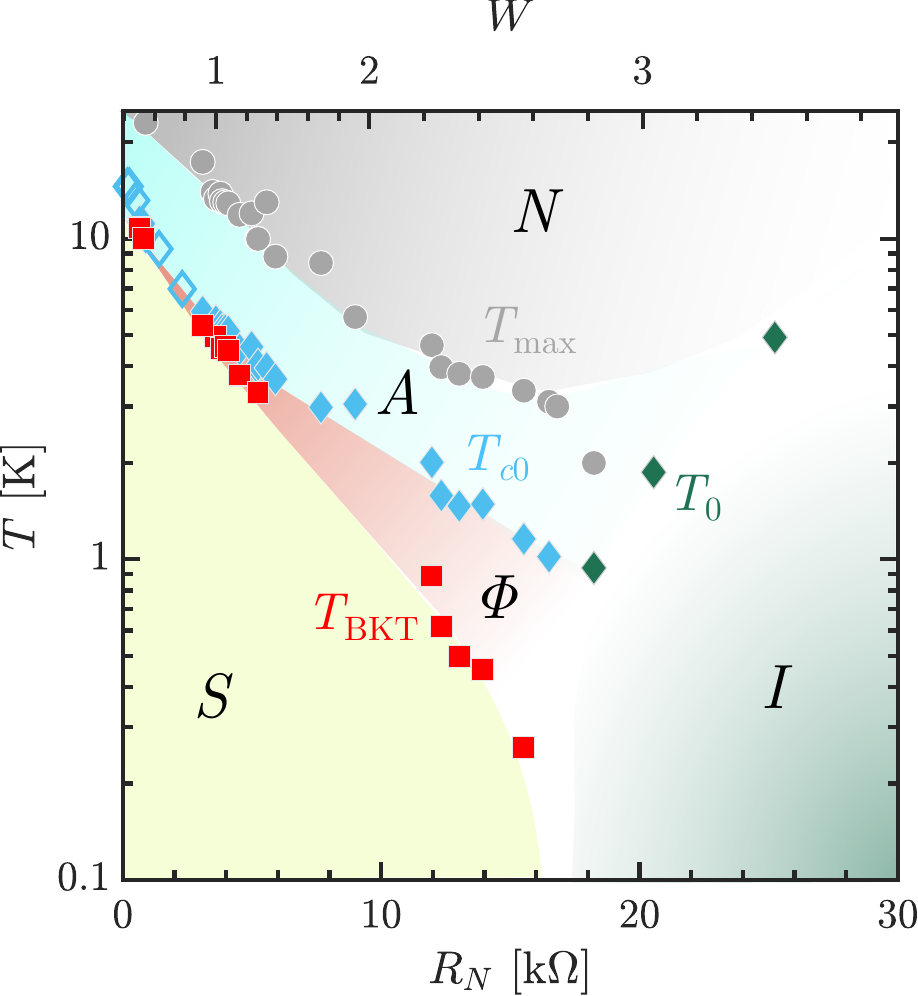}
		\caption{{\bf Phase diagram} Regions in the $(T-\RN)$-plane separated by the BKT phase transition line $\Tbkt(\RN)$, the line  $\Tczero(\RN)$, which denotes the crossover between the phase and amplitude fluctuations regimes, and $T_\text{max}(\RN)$ signaling the crossover to weakly localized metal. The green symbols indicate the activation temperatures in the insulating phase. 
		}
		\label{fig:phasediagram}
	\end{figure}

	%
	\textbf{Summary and Discussion:} \\
	Our results are summarized in Fig.~\ref{fig:phasediagram}, where we display the regions I-IV in the $T$-$\RN$ plane. A zero resistance state exists only in the bottom left region (green) and is separated by the BKT phase transition line (red dots) from the regime of thermal phase fluctuation (red region). At $\Tczero(\RN)$ (blue dots), the phase fluctuation regime crosses over to the amplitude fluctuation regime (light blue region), which upon further increase of temperature eventually approaches the normal state (grey region) at $T_\text{max}(\RN)$ (grey dots). In the bottom left corner, an insulating phase is located (brown region). 
	The degree of insulation is characterized by activation temperatures $T_0(\RN)$ (brown dots) that govern the very large activated resistance
	$\Rsq(T)\propto\exp(T_0/T)$ 
	\cite{Baturina2007-du,Sacepe2015-pp}.   
	
	The SIT is expected to take place at zero temperature where the green and brown regions meet. Towards the SIT, the stiffness at the superconducting phase boundary, 
	following $\Tbkt$, tends to vanish (see also Fig.~\ref{fig:analysis}). In contrast, the pairing  temperature, $\Tczero$, and therefore the order parameter 
	remain finite. Taken together, these experimental results strongly suggest that the SIT is a quantum phase transition driven by quantum  fluctuations of the superconducting phase and, therefore, of second order.

	The overall structure of Fig.~\ref{fig:phasediagram} resembles the schematic form conjectured in \cite{Charpentier2025} with the crucial difference that the conjectured diagram assigns the SIT to be of first order. 
	We consider it likely that the apparent friction between our results and those of~\cite{Charpentier2025} reflects different material systems measured in experiments that have been motivating the respective phase diagrams. 
	The focus of this work was on two-dimensional NbN-films thickness $d_\mathrm{NbN}\simeq 3\,$nm is comparable to the superconducting coherence length $\xi$ with a metallic electron density ($n\simeq 4\cdot10^{22}\mathrm{cm^{-3}}$) and a nano-crystalline structure.     
	In contrast, 
	Charpentier et al. have measured thicker InO$_x$ films ($d_\mathrm{InOx}\simeq 30\,\mathrm{nm} \gg \xi$) that are believed to exhibit three-dimensional behavior with a much lower carrier concentration and exhibit an amorphous structure. 
	The latter suggest that the InO$_x$ samples are situated at much stronger disorder and at significantly weaker screening. 
	
	Indeed, a strong disorder, glass-type scenario has been suggested by Charpentier et al. to explain their experiments \cite{Charpentier2025}. In contrast to our data in Fig.~3f~(inset), they observe a sharp jump to zero of $\Js(0)$, which reflects a discontinuous, i.e., first order transition from the superconductor to a different ground state. 
	The explanation of the emergence of the first order transition that was suggested by Poboiko and Feigelman {\cite{Poboiko2024-rm}} is based on an analysis according to which the glassy state can have a higher entropy 
	than the superconductor - vanishing with $T^3$ as compared to $\exp(-\Delta/T)$, at least on mean-field level~\cite{Poboiko2024-rm}.

	For both materials, the reentrant $R(T)$ indicates that the scaling is eventually cut off.
	As seen in the inset of Fig.~\ref{fig:analysis}f, $\Js(\RN)$ is much flatter (with a much smaller exponent) for InO$_x$ than for NbN, suggesting that this transition occurs at some distance to the quantum critical point at $W_c$. In the case of NbN, the pronounced downturn of $\Js(\RN)$ signals a much more important effect of quantum phase fluctuations prior to the insulating state.\\[2mm]
	\FloatBarrier

	\noindent{\bf Acknowledgements}\\
	We thank M.~Aprili, T. Baturina, J.~Estève, M.~Feigelman, V.~Geshkenbein, I.~Gornyi, A.~Mirlin, P.~Raychaudhuri, B.~Sacepe, and A.~Varlamov for valuable discussions. Two magnetron-sputtered NbN films on the low resistance side were kindly provided by A.~Dutta and P.~Raychaudhuri (TIFR Mumbai). Work in Jena and Regensburg was supported by the European Union’s Horizon 2020 Research and Innovation Program under Grant Agreement No. 862660 QUANTUM E-LEAPS and the Federal Ministry of Science and Technology (BMBF) through Project No.~13N17122 NbNanoQ. I.M. acknowledges financial support by the SNSF via the Swiss Postdoctoral Fellowship Grant No.~TMPFP2 217204. The work was funded partially by the German Research Foundation  (DFG)—Project-ID 314695032–SFB 1277 (Subproject  B08) and projects No.~EV30/11-1, No.~EV30/12-1, No.~EV30/14-1. \\[2mm]
	
	\noindent{\bf Author contributions}\\ A.W. and L.P. fabricated the devices and performed the measurements. A.W., L.P., S.R.~and N.P.~developed and optimized the measurement method. S.L.~and E.I.~conducted the ALD deposition and performed initial characterization of the NbN films. A.W., I.M, and C.S. analysed the data. C.S.~and A.W.~conceived the experiment. A.P., A.D. and F.E. formulated the theoretical model. All authors contributed to discussions and to the writing of the paper.\\

	\bibliographystyle{naturemag}


	\noindent\textbf{\large Methods}\\
	\noindent\textbf{Samples and control of the disorder}\\
	We investigate ALD grown NbN films \cite{Linzen_2017} with a thickness $d = 3.5\pm 0.3$ nm on top of an  550 nm thick amorphous SiO$_2$  (thermally oxidized silicon wafer with [100] orientation). One film film with lower disorder ($\RN\simeq880\,\Omega$) was prepared by magnetron sputtering on a MgO substrate and is similar to the films presented in Ref.~\cite{Mondal_2011b}. For the ALD-grown films, the normal state resistance can be controlled in two ways: (i) variation of the $N_2$ partial pressure during growth, and (ii) gradual oxidation at ambient conditions after growth. We use both approaches as coarse- and fine-tuning knob where $\Rsq(300\,\text{K})$ serves as rough guidance, varying between 1.7 and 3.5\,k$\Omega$ for the least and most disordered samples, respectively. We associate $\Rsq(300\,\text{K})$ with the classical Drude resistance. To enhance sensitivity, we use standard electron beam lithography and ion etching techniques to prepare long ($\sim$ 40-200 squares) meander structures of widths 10 - 200~\textmu m (Figs.~\ref{fig:}a,b).\\[1mm]
	\noindent{\bf Measurements}\\
	Our setup allows for a measurement of AC- and the four-terminal DC-resistance and current-voltage $(IV)$ characteristics of the same device with high sensitivity in the same cooldown. In order to ensure that resistance values are obtained within the linear regime, each point in the $\Rsq(T)$ curves was extracted from a full $I(V)$-curve. This is most essential for low resistance values, as current-induced unbinding of vortex-antivortex pairs becomes more and more easy. 
	
	The kinetic inductance is measured by  integrating our devices into a cold RLC circuit, that permits also access to the DC-resistance.  We measure the transmitted power spectra of the circuit in Fig.~\ref{fig:}c using a vector network analyzer. The resonance frequency is extracted from the sheet kinetic inductance $L_\square$ of the sample \cite{Baumgartner_2020} and typically varies between 0.2\,-\,4\,MHz, depending on temperature.
	
	The sharpness of the BKT transition apparent from $L_\text{kin}(T)$ is caused most probably the smallness of our devices, extending over a small fraction of a mm, as opposed to $\sim10\,\text{mm}$ typically required for the two-coil method \cite{Yong2013,Mondal_2011b,Mandal2020}. A comparison of both methods on the very same NbN films shows that substantial broadening occurs for the two-coil method only. Patterning  small meanders at different positions on the films used for two-coil experiments reveals that both $\Js(0)$ and $\RN$ vary about 10\% between the center and the edge regions of the film [cite A. Weitzel, A. Dutta, L.Pfaffinger et al., in preparation]. As the edges contribute substantially to the signal for the two-coil measurement, such edge inhomogeneity is sufficient to produce the broadening of the BKT transition discussed in \cite{Benfatto2009}.\\[1mm]
	\noindent{\bf Analysis of $\Tczero$}\\
	In Figure~\ref{fig:}e we plot the obtained values of $\Tczero(\RN)$. We expect that the pairing temperature $\Tczero$ is reduced by Gaussian fluctuation: $\Tczero=T_{c00}(1-2Gi |\log Gi|)$ (blue dashed line in Fig.~\ref{fig:}e) \cite{Larkin2005,Koenig2015}, where  $Gi(\RN)=\RN\,7\zeta(3)\,4e^2 /(\pi^3h)=1.086\,\RN e^2/h$ is the Ginzburg-Levanyuk number \cite{Koenig2015}.   
	A second, even more important effect originates from the suppression of the screening of Coulomb interaction due to the strong disorder \cite{Finkelstein1994-ub}.  It reads
	\begin{align}
		\frac{T_{c0}}{T_\mathrm{c00}}=\exp\left(-\frac{1}{\gamma_c}\right)\left(\frac{\gamma_c+\sqrt{t_D}/2}{\gamma_c-\sqrt{t_D}/2}\right)^{1/\sqrt{t_D}}
		\label{eq:finkelstein}
	\end{align}
	where $T_\mathrm{c00}$ is the pairing temperature in the clean case, $\gamma_c=ln[\hbar/(k_BT_{c00}\tau)]$ with mean free time $\tau$ and $t_D=\frac{\pi}{2}\frac{e^2}{h}R_{Drude}$ proportional to the unrenormalized Drude sheet resistance $R_{Drude}$. Literature data for thick, ALD-deposited NbN films with $\RN\ll h/e^2$ \cite{Linzen_2017}, suggest a lower disorder limit of $T_{c00}=14$~K, which is slightly lower than that of bulk NbN (17\,K). 
	As shown in ExtendedDataFig.~\ref{fig:ext_data:finkelstein}, the combination of the Gaussian fluctuations with the Finkel'stein comtribution (light blue line) describes the the measured $\Tczero(\RN)$ very satisfactorily (dark blue line) with only one free parameter $\gamma_c=5.312$.\\
	
	\noindent{\bf Theoretical Methods}\\
	All results of mean-field calculations have been produced with the computational methods explained in detail in Ref.~\cite{Stosiek2020}. In short, we have considered an attractive-U (spin-full) Hubbard model on the square lattice in two dimensions within the mean-field (BdG-type) approximation. The system used is of size 96 x 96 at particle density $n=0.875$. Disorder has been introduced by uncorrelated on-site potentials $V_i$, that have been drawn from a box distribution $[-W,W]$. Units are such that the bandwidth of the disorder-free (clean), non-interacting sample reaches from $[-4,4]$. 

	The self-consistency loops have converged quasi-particle energies and wavefunctions. The iteration cycle was stopped when the charge density converged to an accuracy better than $1\%$. \\[2mm]
	\noindent{\bf Mapping between $\RN$ and $W$ }\\
	A reliable link is established between the level of disorder measured by sheet resistance $\RN$ in the normal state and the strength $W$ of the Anderson disorder. It relies on the fact that these observables determine $t_{c0}=(\Tczero-\Tbkt)/\Tczero$ in experiment and theory, respectively. Eliminating $t_{c0}$ from the two equations $t_{c0}(\RN)$ and $t_{c0}(W)$ establishes the relation $W(\RN)$ (Fig.~\ref{fig:sup:R_vs_W}) and thus allows a direct mapping between experiment and theory in the different regimes of disorder. To compare numerical $J_s(W)$, $E_g(W)$ and $E^*(W)$, which are given in units of the hopping integral $t$, to their experimental counterparts, given in units Kelvin, raw numerical data are multiplied by a conversion factor. This factor is $42.0$ ($113.4$) $\mathrm{K}/t$ for $J_s$ ($E_g(W)$, $E^*(W)$) in Fig. \ref{fig:analysis} f (g).
	
	\FloatBarrier
	
	
	\beginextendeddata
	\begin{figure*}[htbp]
		\includegraphics[width=\textwidth]{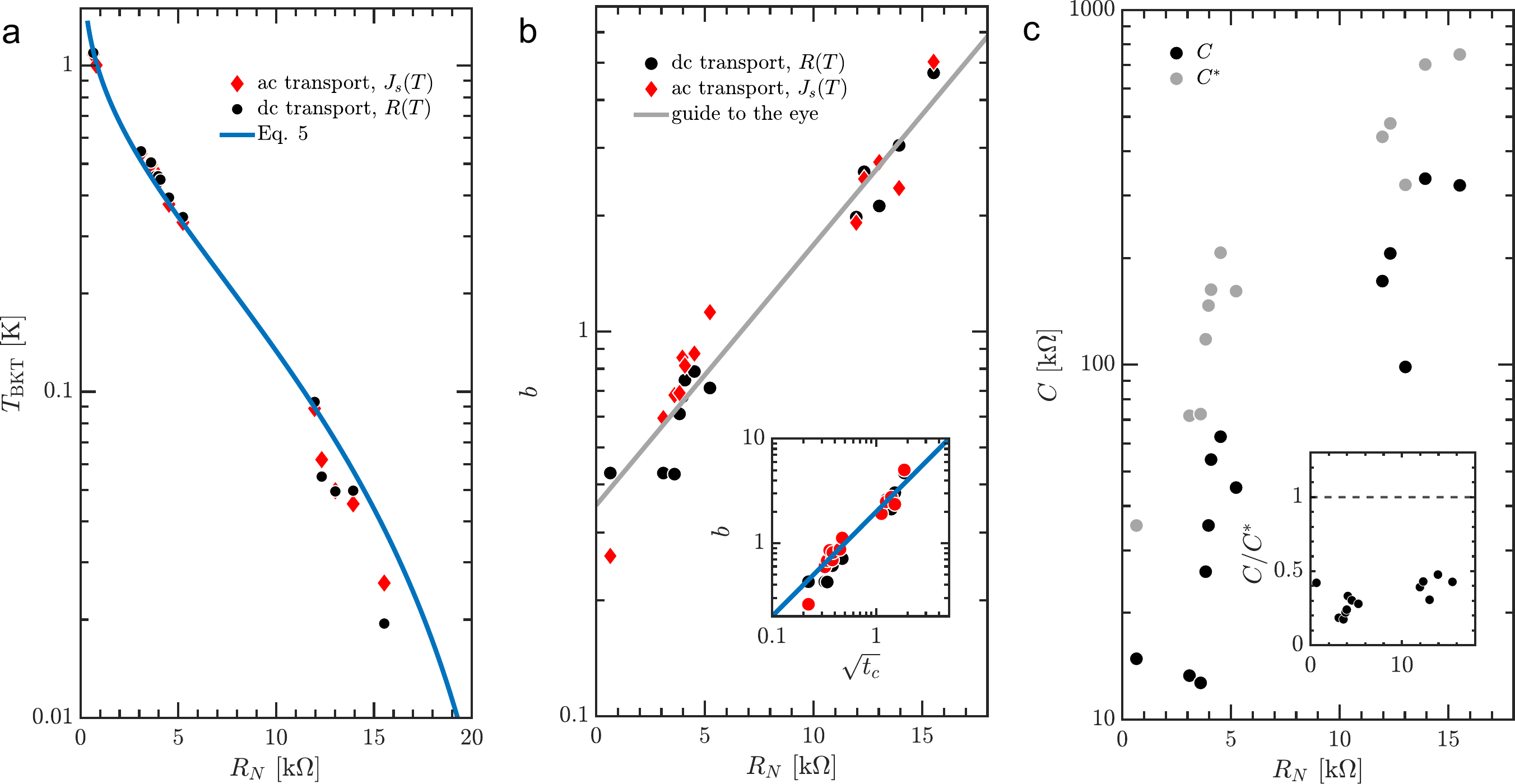}
		
		\caption{{\bf Evolution of the fitting parameters of Eq.~\ref{eq:SRC} with normal-state resistance $\Rsq$: a)} Black dots: Vortex unbinding temperature $\Tbkt$, derived from the three parameter fit of $R(T)$ according to Eq.~\ref{eq:SRC} below the inflection point. Red dots: $\Tbkt$ derived from $J_s(T)$ through the requirement $J_s(\Tbkt)=2\Tbkt/\pi$. Blue line: Varlamov-Larkin prediction (Eq. 5), based on the Finkel'stein fit of $\Tczero$.  {\bf b)} Black dots: parameter $b$ of Eq. \ref{eq:SRC}. Red dots: $b$ based on experimentally determined phase stiffness $J_s$, vortex excitation energy $\mu$, $\Tczero$ and $\Tbkt$ according to $	b\simeq \frac{4}{\pi^2}\frac{\mu(0)}{J_s(0)}\sqrt{\frac{T_{c0}-T_\mathrm{BKT}}{T_\mathrm{BKT}}}$. Grey line: empricial fit according to $b_0\exp(R_N/R_{b0})$, with $b_0= 0.3533$, $R_{b0}=6.4144$ k$\Omega$. Inset: $b$ vs reduced temperature $\sqrt{t_c}=\sqrt{\frac{\Tczero-\Tbkt}{\Tczero}}$. Blue line:  $b(t_c)=2\sqrt{t_c}$, expected within the $XY$ model. {\bf c)} Black dots: Prefactor $C$ of Eq. \ref{eq:SRC}. Grey dots: Theoretically expected value $C^*$ based on normal-state resistance, $t_c=(\Tczero-\Tbkt)/\Tczero$ and the third parameter $b$ (\cite{Hebard1983-cw, HalperinNelson_1979}). Inset: ratio $C/C^*$, clustering around 1/3.
		}
		\label{fig:ext_data:bkt_parameters}
	\end{figure*}

	\begin{figure*}[htbp]
		\includegraphics[width=\textwidth]{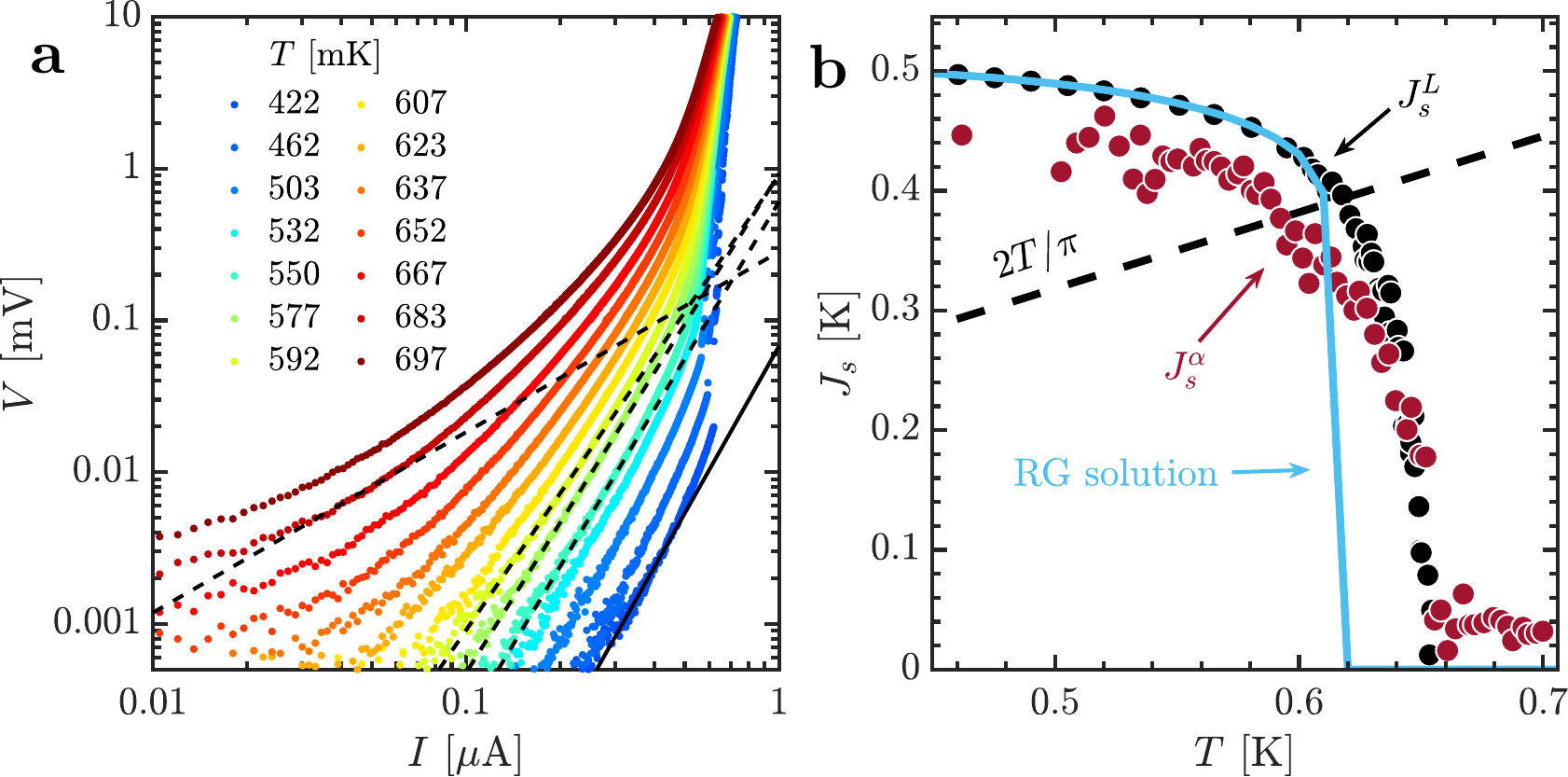}
		
		\caption{{\bf Non-linear transport near the BKT transition at high disorder a)} $V(I)$ traces at various temperatures obtained by pulsed measurement ($t_{meas}\sim1$ s, to minimize Joule-heating of the sample stage via the decoupling resistors of the resonator) of a film with $R_N=12.3$ k$\Omega$ close to the SIT at $R_N\sim16$ k$\Omega$. Dashed lines: Least-square fits of the low-power regime of logarithmically scaled data to a power-law model $\log(V)=\alpha\times\log(I)+A$. {\bf b}: Superfluid stiffness $J_s^L$ obtained by kinetic inductance (black) and $J_s^\alpha$ (dark red) from power-law fits in the left panel. Solid, blue: RG fit to black dots, discussed above. Dashed black: Universal HN line $J_s=2T/\pi$.}

		\label{fig:ext_data:non_linear_transport}
	\end{figure*}

	\begin{figure*}[htbp]
		\includegraphics[width=0.8\textwidth]{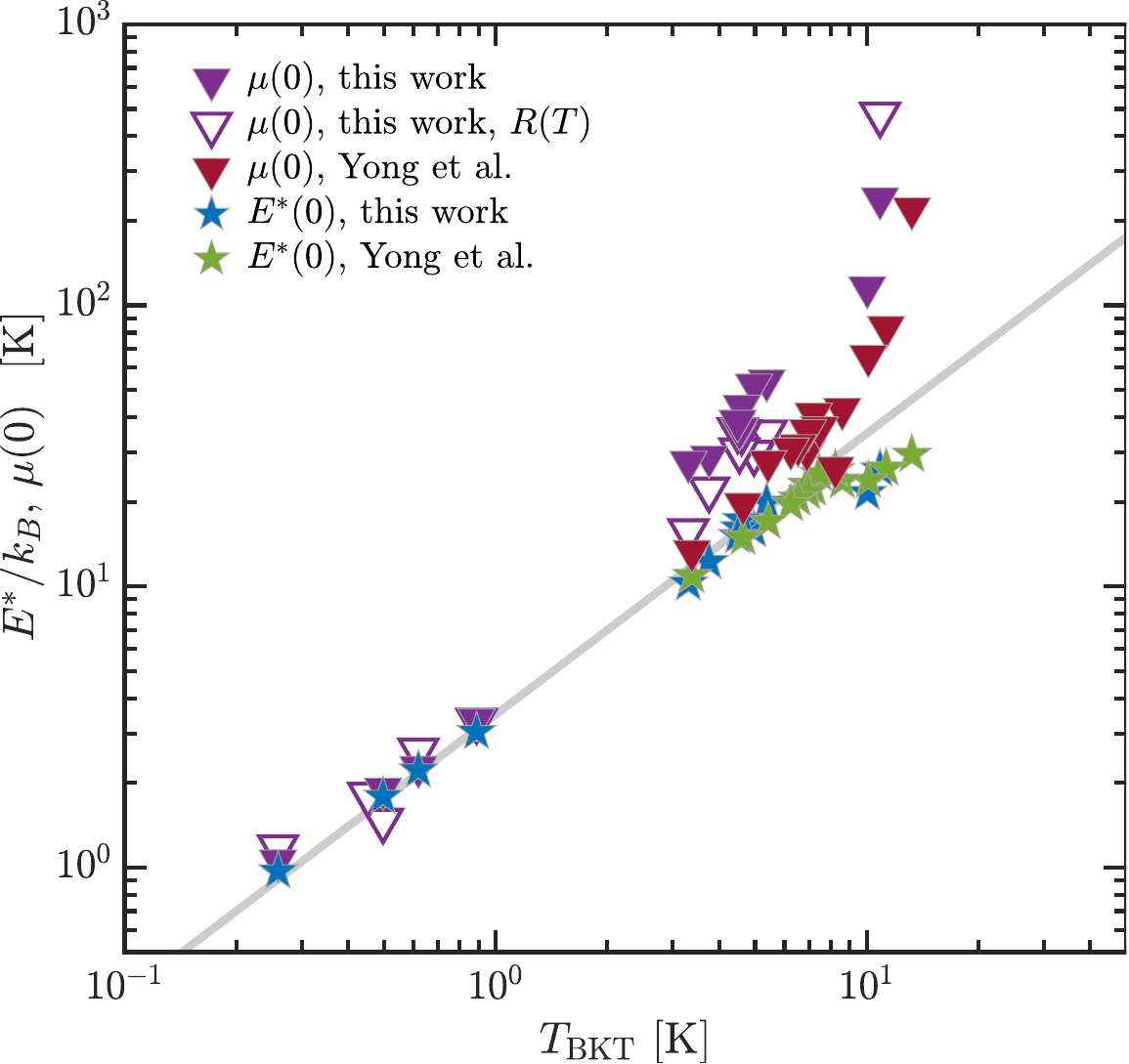}
		
		\caption{{\bf Characteristic temperatures $\mu$ and $E^*$ of thin NbN films compared to \cite{Yong2013} } Values of vortex-antivortex pair excitation energy $\mu$ (triangles) and thermal activation energy $E^*$ (stars) as a function of vortex unbinding temperature $T_\mathrm{ BKT}$ of thin NbN films, obtained in this work (purple and blue) and obtained in previous two-coil experiments by \cite{Yong2013} (red and green). The grey line is a guide to the eye with slope 1.}
		\label{fig:ext_data:yong_vs_weitzel}
	\end{figure*}

	\begin{figure*}[htbp]
		\includegraphics[width=\textwidth]{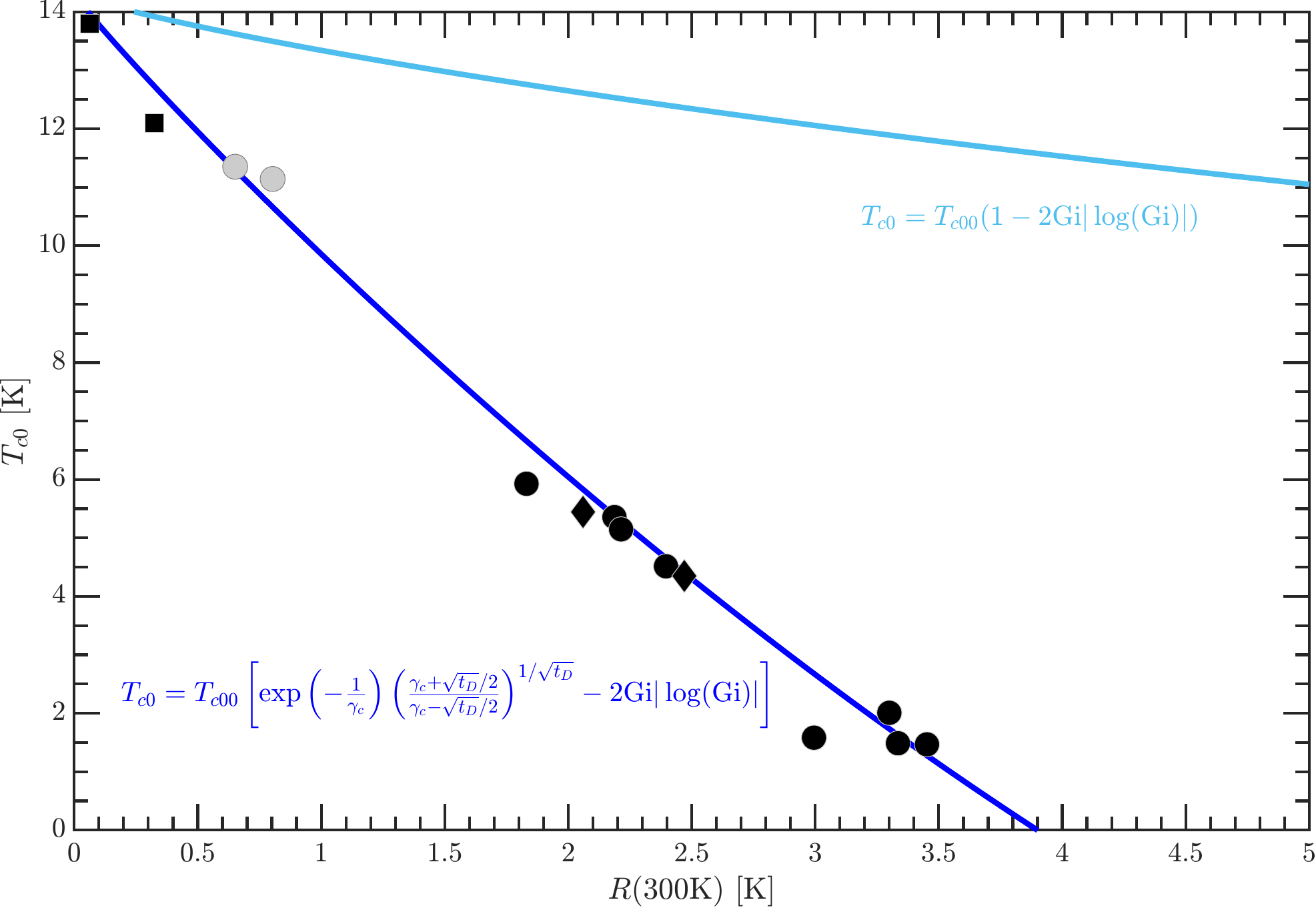}
		
		\caption[Fermionic suppression of $T_{c0}$]{\textbf{Fermionic suppression of $T_{c0}$:} Mean field critical temperature  $T_{c0}$ as a function of normal state sheet resistance at room temperature. Light blue: Suppression of $T_{c00}=14.3$ K \cite{Linzen_2017} by Gaussian amplitude fluctuations. Dark blue: Combined suppression of Finkel'stein suppression (Eq. \ref{eq:finkelstein} with single adjustable parameter $\gamma=5.312$) and amplitude fluctuations. Black dots: ALD deposited samples examined in this work. Black rectangles: data from ALD deposited films (\cite{Linzen_2017}). Black diamonds with no corresponding measurement of $\Js$. Grey dots: data from magnetron sputtered NbN films  provided by A.~Dutta and P.~Rychaudhuri (Tata Institute for Fundamental research, Mumbai). The values agree well with those in Ref.~\cite{Mondal_2011b}.} 
		\label{fig:ext_data:finkelstein}
	\end{figure*}

	
	\stopextendeddata
	
	\FloatBarrier
	
	\beginsupplement
	\preprint{APS}
	
	\title{Supplementary Information for "Origin of the superconductor-insulator transition in disordered two-dimensional films"}
	
	\author{Alexander Weitzel,$^1$ Lea Pfaffinger,$^1$ Animesh Panda,$^2$ Ilaria Maccari,$^3$ Simon Reinhart,$^{1*}$ Sven Linzen,$^4$ Evgenii Il'ichev,$^4$ Nicola Paradiso,$^1$ Ferdinand Evers,$^{2,5}$  Christoph Strunk$^{1,5}$}
	\affiliation{$^1$Institute for Experimental and Applied Physics, University of Regensburg, 93040 Regensburg, Germany}
	\affiliation{$^2$Institute for Theoretical Physics, University of Regensburg, 93040 Regensburg, Germany}
	\affiliation{$^3$Institute for Theoretical Physics, ETH Zurich, 8093 Zurich, Switzerland}
	\altaffiliation{School of Applied and Engineering Physics, Cornell University, Ithaca, NY, 14853, USA}
	\affiliation{$^4$ Leibniz Institute of Photonic Technology, 07745 Jena, Germany}
	\affiliation{$^5$ Halle-Berlin-Regensburg Cluster of Excellence CCE, University of Regensburg, 93040 Regensburg, Germany}
	
	\date{\today}
	
	\onecolumngrid
	
	\begin{center}
		\textbf{\Large Supplementary Information} 
		
		\vspace{0.5cm}
		
		\noindent \textbf{\large Origin of the superconductor-insulator transition in disordered two-dimensional films}
	\end{center}

	\section{Materials and Methods}
	\subsection{NbN wafers}
	We fabricate our devices from 3 nm thin ALD grown NbN films on an amorphous, about 550 nm thick SiO$_x$ layer onto silicon substrate (thermally oxidized silicon on [100] orientated wafers).  The samples discussed in the main text originate from six different wafers, assigned a running number (327, 334, 388, 397, 400, 513).  During deposition, the disorder of the films is coarsely determined by the number of deposited layers and the nitrogen stoichiometry.
	After deposition, the disorder is gradually increased by aging at ambient conditions. Thus, the normal state resistance of wafer 397 increased by roughly 20\% over the course of one year. The aging process is accelerated at elevated temperatures ($\sim100^\circ$C). The susceptibility of the material towards this aging process is higher for more resistive films. Less disordered wafers (e.g.~wafer \#400) showed excellent reproducibility of results if different devices made from the same wafer were measured in short succession.\\
	Additionally, two films were fabricated from magnetron sputtered NbN films with properties similar to those in Ref.~\cite{Mondal_2011b} on MgO substrate.

	\subsection{Measurement techniques}
	\paragraph{Linear dc transport above $T_\mathrm{BKT}$}:
	
	$R(T)$ is measured in four-contact geometry using a voltage source (Yokogawa GS200) and a
	large preresistor $R_p = 100~\mathrm{M\Omega}$, such that the current is well approximated by
	$I \approx V/R_p$ to within 1\%. In a second configuration, the current is measured directly
	via a transimpedance preamplifier (Femto DDPCA-300-S) in reverse mode, which was the more
	commonly used setup in this work. The voltage drop across the sample is measured with a
	multimeter (Agilent 34410A, 3458A, or 34470A), preceded by a voltage preamplifier (Femto
	DLPVA) with variable gain up to 20000~V/V, used at maximum gain close to $T_\mathrm{BKT}$
	where the sample resistance is very small. For dc measurements, all measurement lines are filtered by a
	$\pi$-filter with cutoff frequency 10~MHz. Each data point corresponds to a $V(I)$
	characteristic from which the zero-bias resistance is extracted by a linear fit.
	
	\paragraph{Non-linear dc transport below $T_\mathrm{BKT}$}:
	
	Non-linear $V(I)$ characteristics below $T_\mathrm{BKT}$ require sweeping the current over several orders of magnitude, ideally within a short time interval to minimize Joule heating via the decoupling resistors $R_D$. In both setups used, the voltage drop is measured simultaneously at maximum and zero preamplifier gain, enabling continuous sweeps covering seven orders of magnitude in $V$ with an overlap of approximately one decade, over which agreement between the two datasets is carefully verified. In the first setup, a second voltmeter is added in parallel to the configuration described above, with data stored in device memory to allow fast voltage ramp programs. In the second setup, the voltage source and voltmeters are replaced by a Nanonis Tramea FPGA, enabling fast and automated measurement routines; the circuit is otherwise identical.
	
	\paragraph{AC response}: 
	
	$L(T)$ is measured with a vector network analyzer (VNA), i.e., Rohde und Schwarz ZNL3. At the input and output we mount bias tees ZX85 SG12 to suppress any residual dc current. The power is chosen such that a reduction in power does not affect the quality factor of the circuit (see below). We amplify the signal at room temperature by 56 dB with a Miteq amplifier. For most of the measurements of strongly disordered samples, we used a cryogenic amplifier of type CITLF1S by cosmic microwave technologies with gain $\sim 20$ dB at low temperature and input noise temperature $< 7$ K, which is thermalized at the 4 K stage.\\

	\subsection{Circuit Design}
	
	\begin{figure}[t]
		\includegraphics[width=.75\textwidth]{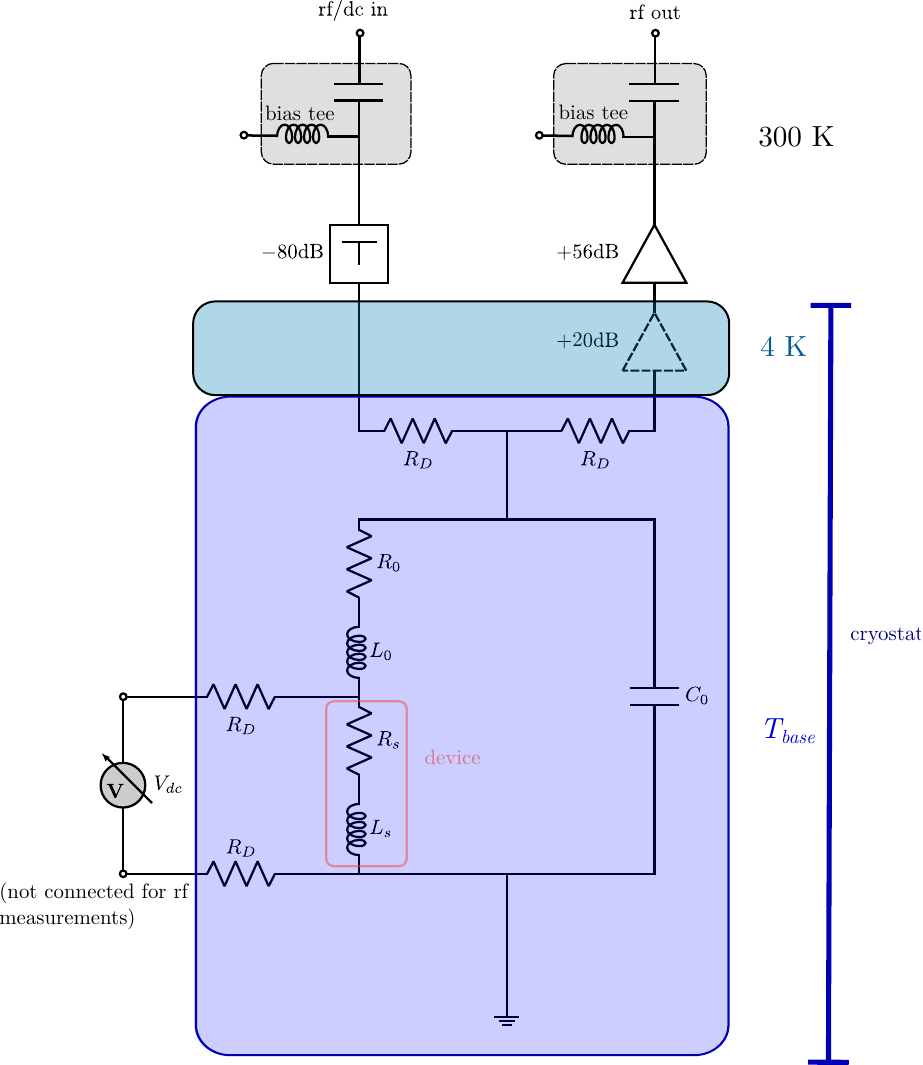}
		\caption[Schematic circuit diagram for simultaneous dc and rf measurements]{\textbf{Schematic circuit diagram for simultaneous dc and rf measurements} The blue area symbolizes the cryostat. The rf excitation is attenuated at room temperature with standard T-attenuators, typically by $-80$ dB. Inside the cryostat, the LC circuit is isolated from the environment by decoupling resistors, $R_D=1-1.4$ k$\Omega$. The circuit itself consists of a SMD capacitor $C_0$, a resistance $R_0$ due to contact resistances and bond wires, a residual inductance $L_0$ due to bond wires and pogo pins of the sample holder and the sample. The sample (red rectangle) contributes $R_s$ to the resistance and $L_s=L_K^{tot}+L_G$ to the inductance, where $L_K^{tot}$ and $L_G$ are the total kinetic and geometric inductance respectively. In this work, $L_K^{tot}\gg L_G$, such that $L_s\approx L_K^{tot}$. The transmitted signal is amplified in particular cases at 4 K and in all cases at room temperature. Via dc contacts, additional dc experiments are possible in the same experimental run.}
		\label{fig:sup:RLCcircuit}
	\end{figure}
	The sample under study is placed in a cold RLC circuit, shown in Fig.~\ref{fig:sup:RLCcircuit}. $L_0$ and $R_0$ denote the residual inductance and resistance arising from bond wires, contact resistances, and a dedicated copper coil ($L_0 \sim 200$ nH); these elements are connected in series with the sample and in parallel with the SMD capacitor $C_0$. Four resistors $R_D$ are included to decouple the resonator from the environment, i.e., from the lead impedances.
	To accommodate the varying geometric constraints of different cryostat sample spaces, a separate resonator assembly is fabricated for each cryostat.
	These realizations differ slightly in the values of $L_0$ and $C_0$.
	In all cases, an unloaded resonance frequency of $3$--$5$ MHz is achieved by appropriate choice of $L_0 \sim 200$ nH and $C_0 \sim 4$--$10$ nF.
	In the He$^4$ setup, the very large kinetic inductance of the long meander structures ($L_K^\mathrm{tot} \sim 200$ nH) made it possible to omit the copper coil entirely.
	Measurements of comparable samples across different resonator realizations show excellent agreement, indicating that the specific circuit implementation has negligible influence on the results.


	
	
	%
	
\begin{figure}[h!]
		\includegraphics[width=\textwidth]{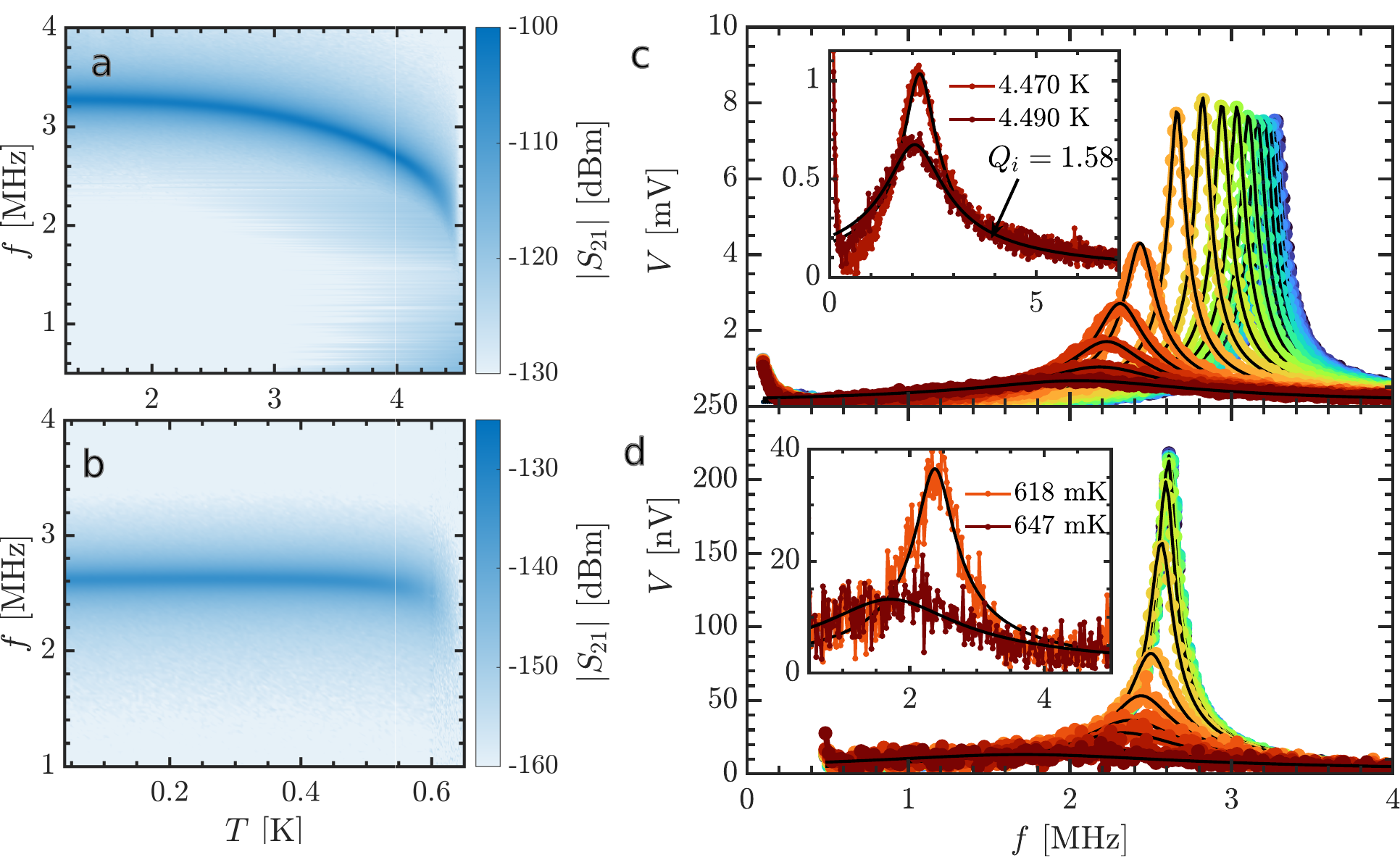}
		\caption[Exemplary rf datasets]{\textbf{Exemplary rf datasets} (a), (b): Amplitude of the complex transmission matrix element $|S_{21}|$ for two devices with $T_\mathrm{ BKT}=4.5$ and $0.6$ K, respectively as function of frequency and temperature. $|S_{21}|$ is given as bare value without amplification. The strongly differing values on the $z$-axes (colorbars) reflect the difference in excitation (see next section). Dark regions, where $|S_{21}|$ is maximal, indicate resonance. (c),(d): Exemplary linecuts in (a), (b) at constant temperature. Blue and red colors correspond to lower and higher $T$, respectively. Solid lines are fits to Eq. \ref{eq:sup:V^2(f)}, yielding $f_0$. Insets: Zoom-in to the critical region where $Q_i$ is strongly suppressed through thermally excited vortices. A lineshape consistent with Eq. \ref{eq:sup:V^2(f)} can still be observed. }
		\label{fig:supp:resonator_data}
	\end{figure}

	
	Using the transmission matrix formalism, $S_{21}$ can be calculated for a network of three impedances, where one is connected to ground. The latter refers to the impedance of the inner circuit, while the others are the decouple resistors. The resulting transmission component reads
	\begin{equation}
		S_{21}(\omega)= \frac{2Z_lQ}{C_0R_x^2}\frac{1}{\omega_0+2iQ{(\omega-\omega_0)}}
	\end{equation}
	Therefore, the measured signal can be described by
	\begin{equation}
		V^2(f)=A\cdot\bigg|\frac{Z_lQ}{\pi C_0R_x^2}\frac{1}{f_0+2iQ({f-f_0})}\bigg|^2
		\label{eq:sup:V^2(f)}
	\end{equation}
	where $V$ is the voltage at the input of the network analyzer, $f$ is the frequency, $A$ is a scaling parameter, Q is the quality factor, $R_x=R_D+Z_l$ with $R_D=992\,\Omega$ being the decoupling resistance and $Z_l=50\Omega$ is the impedance of the cables.
	Changes of resonance frequency can be directly connected to changes of the sample's kinetic inductance.\\
	Fig. \ref{fig:supp:resonator_data} shows exemplary datasets obtained for a sample with $R_N=4.1$ k$\Omega$ (upper row) and a second sample with $R_N=12.3$ k$\Omega$ (lower row). 
	
	\subsection{Optimization of rf Measurement Conditions}
	
\begin{figure}[htbp]
		\includegraphics[width=.75\textwidth]{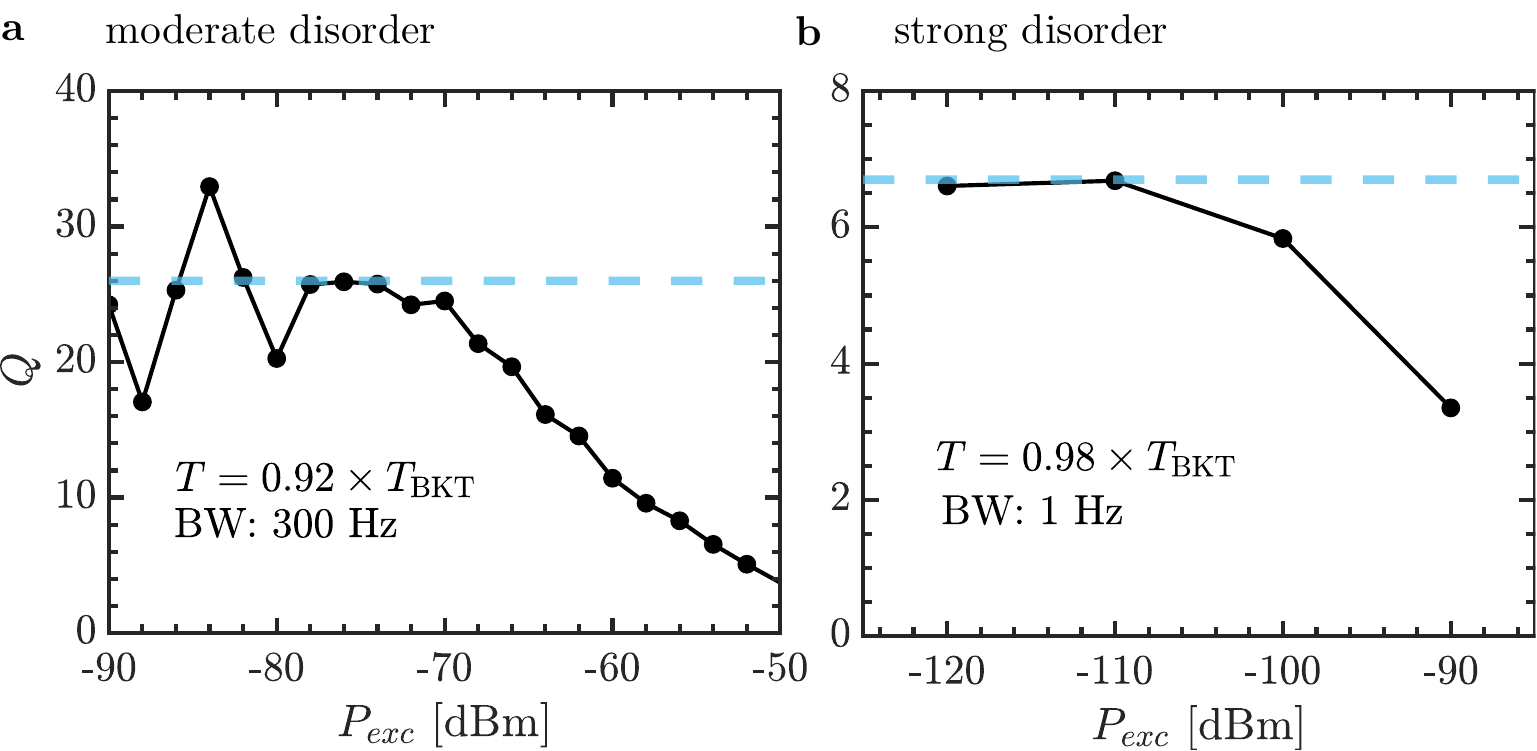}
		\caption[Quality factor as function of drive power]{\textbf{Quality factor as function of drive power}. $Q(P_{exc})$ is shown for two exemplary samples in (a), (b). $(P_{exc}$ is the output drive power of the network analyzer after attenuation. Device in (a), (b): $T_\mathrm{ BKT}=5.5$, $0.5$ K, respectively. }
		\label{fig:sup:QofP}
	\end{figure}

	\paragraph{Drive Power}:
	
	At sufficiently large excitation power, the rf drive can induce vortex unbinding in the sample, leading to excess damping below $T_\mathrm{BKT}$. To identify the critical drive power $P_c$ below which $Q(P)$ saturates, $Q$ is measured as a function of drive power prior to each experiment (Fig.~\ref{fig:sup:QofP}). For films with $T_\mathrm{BKT} \approx 4$~K measured in a He$^4$ setup without cold preamplification, a drive power of $-80$~dBm at a bandwidth of $10$--$100$~Hz was found to be sufficient. For samples close to the SIT with $T_\mathrm{BKT} < 1$~K, measured in a dilution refrigerator with a cold preamplifier at the 4~K stage, the drive power was reduced to $-110$~dBm at a bandwidth of $1$~Hz, reflecting the enhanced sensitivity of the weakly superconducting state to external perturbation.
	
	\paragraph{Compensation of Stray Magnetic Field }:
	\begin{figure}[htbp]
		\includegraphics[width=.5\textwidth]{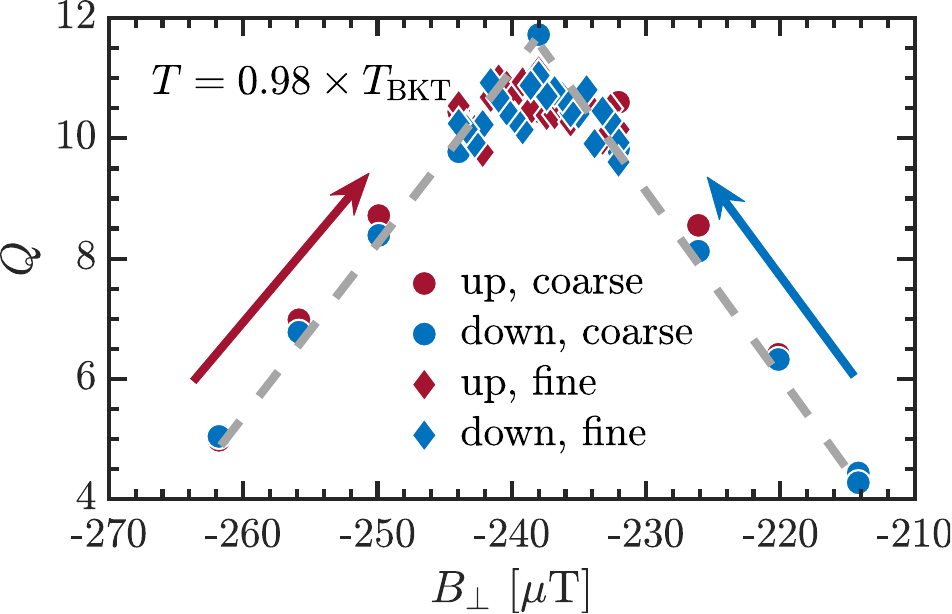}
		\caption[Quality factor of as function of magnetic field]{\textbf{Quality factor of as function of magnetic field}. Shown is $Q(B)$ for a representative device with $T_\mathrm{ BKT}=4.5$ K. Solid lines are guides to the eye. Arrows indicate sweep direction. Previously published in and slightly adapted from Supplement of (\cite{Weitzel2023}).}
		\label{fig:sup:QofB}
	\end{figure}
	
	Stray perpendicular magnetic fields introduce free vortices that are indistinguishable from thermally excited ones, and must therefore be carefully compensated prior to each experiment.
	The high sensitivity of $Q$ to dissipation makes the resonator a precise probe of vortex concentration, particularly close to $T_\mathrm{BKT}$ where the superfluid stiffness $J_s$
	is reduced. The compensation field is determined by sweeping the solenoid field over a small range ($\sim$tens of $\mu$T) and identifying the maximum in $Q(B)$ by linear extrapolation (Fig.~\ref{fig:sup:QofB}). Negligible hysteresis between sweep directions was consistently observed, indicating weak vortex pinning across all investigated films. The field resolution achievable in this procedure scales inversely with meander width; for a 10~$\mu$m wide meander the uncertainty is approximately 12~$\mu$T, comparable to the field equivalent of one flux quantum per square  $B_\mathrm{min} = 21~\mu$T). After the optimum is identified, the sample is heated above $T_{c0}$ to expel residual vortices. Provided the solenoid is cooled from room temperature in a low-field state and subsequently not cycled to large fields, the optimal compensation field was found to remain stable over timescales of months.

\begin{figure}[h]
		\includegraphics[width=.5\textwidth]{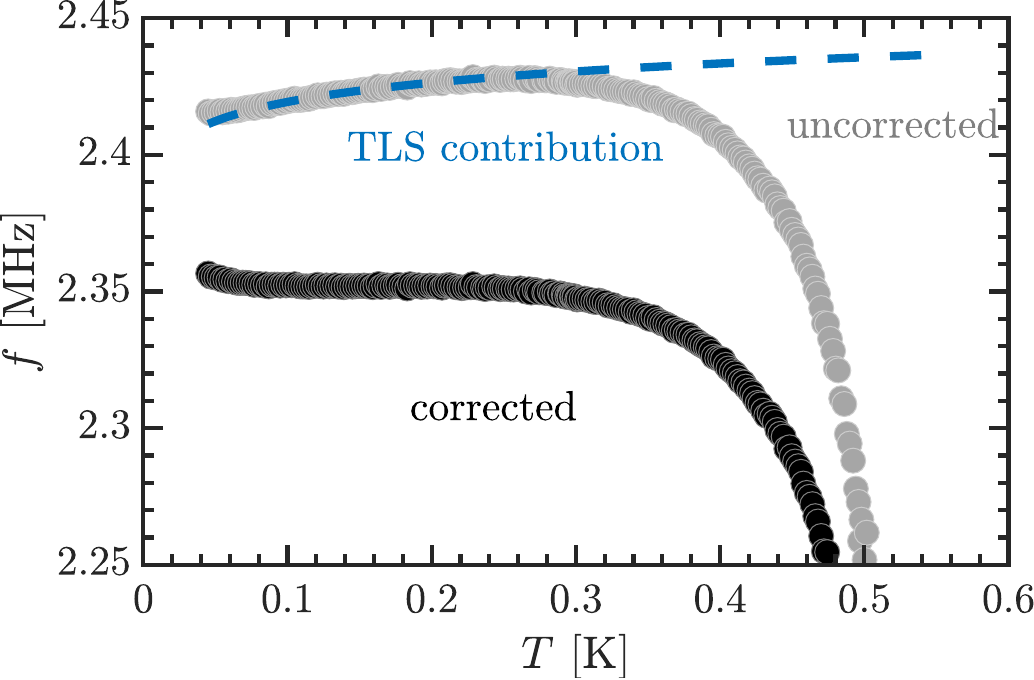}
		\caption[Two-level system contribution to $f_0(T)$]{\textbf{Two-level system contribution to $f_0(T)$}. Shown is $f_0(T)$, as extracted from fits of Eq. \ref{eq:sup:V^2(f)} to $|S_{21}(T,f)|$ for a device close to the SIT with $T_\mathrm{ BKT}=0.5$ K as grey symbols. The saturation for $T<100$ mK is an experimental artifact due to slow device thermalization. Blue dashed line is a fit of Eq. \ref{eq:TLS_f0} to the low-temperature part ($120\text{ mK}<T<200\text{ mK}$) of $f_0(T)$. Black symbols are $f_0(T)$ with the contribution of TLS subtracted.}
		\label{fig:sup:tls}
	\end{figure}

	\paragraph{Two-Level Systems}:
	
	At mK temperatures, the kinetic inductance contribution to $f_0(T)$ can be obscured at low temperatures by two-level systems (TLS) in the SiO$_x$ substrate, which couple to the rf electric field and produce a logarithmic temperature dependence of $f_0$ (Fig.~\ref{fig:sup:tls}). In the weak-driving limit, the TLS-induced shift in resonance frequency is given by~\cite{Pappas2011-hx}:
	
	\begin{align}
		\label{eq:TLS_f0}
		f_0^\mathrm{TLS}(T) = f_0(0) + f_0(0)\frac{F\delta_\mathrm{TLS}^0}{\pi}
		\left[\mathrm{Re}\,\Psi\!\left(\frac{1}{2}+\frac{hf}{2\pi i k_BT}\right)
		-\log\!\left(\frac{hf}{2\pi k_BT}\right)\right]
	\end{align}
	
	where $\Psi$ is the digamma function, $F$ the filling factor, and $\delta_\mathrm{TLS}^0$ the loss tangent of the TLS-hosting medium. Equation~\ref{eq:TLS_f0} is fitted to the low-temperature regime of $f_0(T)$ (typically $120~\mathrm{mK} < T < 200~\mathrm{mK}$), where quasiparticle excitations are still exponentially suppressed, with $f_0(0)$ and $F\delta_\mathrm{TLS}^0$ as free parameters. The TLS contribution is then subtracted, yielding $f_0(T)$ governed by superconducting physics alone. The extracted $F\delta_\mathrm{TLS}$ is corrected by a factor $L_\mathrm{tot}/L_K^\mathrm{tot} \approx 1.4$ to account for energy stored outside the superconductor, yielding $F\delta_\mathrm{TLS} = 0.191$, consistent with previous  tudies on substrates of similar composition~\cite{LorenzDiss}.
	
\begin{figure}[h]
		\includegraphics[width=.5\textwidth]{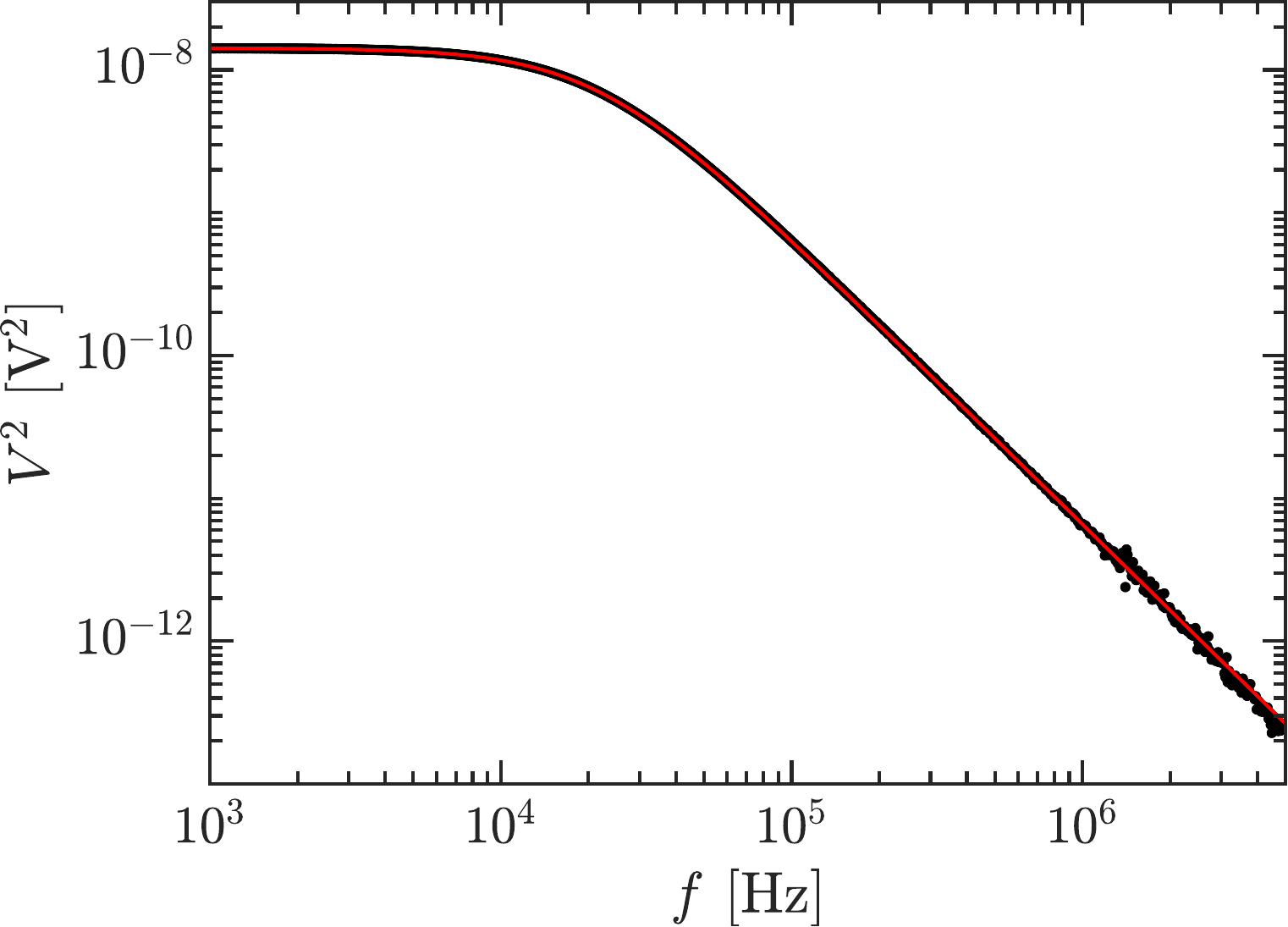}
		\caption[Calibration of capacitance]{\textbf{Calibration of capacitance} Sweeping $f$ over several orders of magnitude at a temperature $T\approx2T_{c0}$, where the sample resistance is $\sim1$ M$\Omega$ reveals low-pass behavior of $|S_{21}|$ (given here in $V^2$). A fit of Eq. \ref{eq:lowpasscalib} (red) to the data yields the capacitance $C$. Previously published in and taken from Supplement of (\cite{Weitzel2023}).  }
		\label{fig:sup:C0}
	\end{figure}

	\paragraph{Calibration of Capacitance}:
	
	The capacitance $C = C_0 + C_\mathrm{stray}$ is determined \textit{in situ} with the sample in the normal state ($T \approx 2T_{c0}$), where the circuit reduces to a low-pass filter formed by $R_D$ and $C$. A frequency sweep over several decades yields $|S_{21}(f)|$, which is fitted by:
	
	\begin{align}
		\label{eq:lowpasscalib}
		V^2(f) = A \cdot \left|\frac{Z_0}{R_x + i\pi C f R_x^2}\right|^2
	\end{align}
	
	with $R_x = R_D + Z_0$ and $Z_0 = 50~\Omega$, from which $C$ is extracted as a fit parameter (Fig.~\ref{fig:sup:C0}). This calibration is performed in every cooldown to account for variations in $C_\mathrm{stray}$. The resulting $C$ is consistently close to the nominal SMD capacitor value $C_0 = 4$--$10$~nF.
	\FloatBarrier
	
	\subsection{Extraction of critical temperatures from dc measurements}
	The mean-field pairing temperature and vortex-unbinding temperature can be extracted from the $R(T)$ data, as discussed in the main text. Here we discuss technical details of this procedure.
	
\begin{figure}[h!]
		\includegraphics[width=0.7\textwidth]{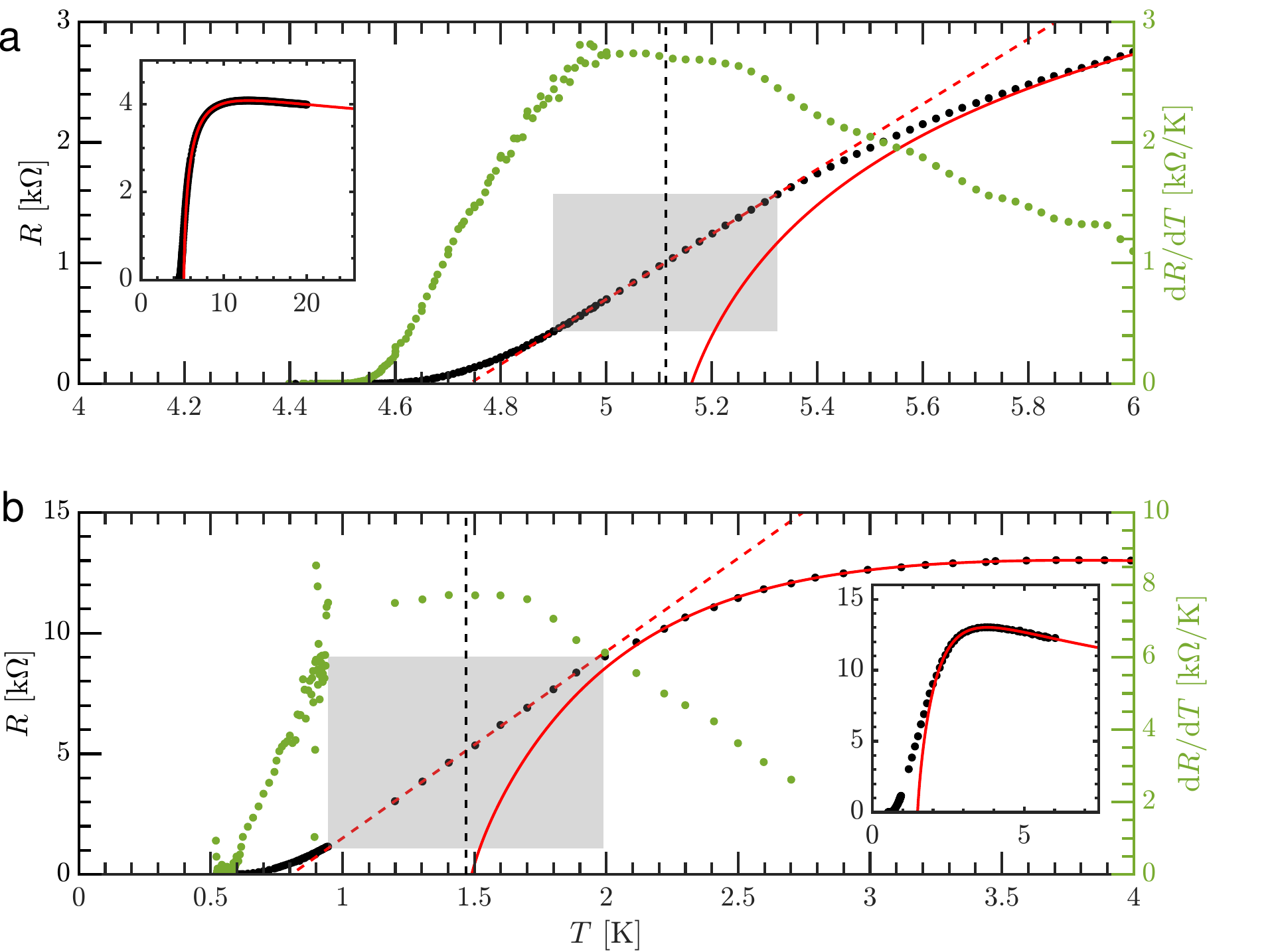}
		\caption[Determination of $T_{c0}$]{\textbf{Determination of $T_{c0}$:} \quad The $\mathrm{d}R/\mathrm{d}T$ curves obtained in experiment always exhibit a peak close to $T_{c0}$, where $R(T)$ has an inflection point. Such a peak can be seen in (a) and (b), where we plot $\mathrm{d}R/\mathrm{d}T$ as green symbols on the right axes. The center of the linear regime in $R(T)$ (flat range in $dR(T)/dT$) agrees well with the value of $\Tczero$ extracted from the fit of amplitude fluctuations (which is determined mainly by the Aslamazov-Larkin contribution). }
		\label{fig:sup:T_c0_analysis}
	\end{figure}

	\paragraph{Mean field pairing temperature $\Tczero$}:
	
	A quantitative theory of the quantum corrections to the conductivity due to interference effects and due to due to superconducting fluctuations is given in Ref. \cite{Glatz2011-af}. Close to $\Tczero$, approximate expressions capture the experimentally observed behavior well \cite{Baturina_2012}. As demonstrated in \cite{Weitzel2023}, strongly disordered NbN films are well described by these approximate expressions, with three free parameters. In \cite{Weitzel2023}, a sanity check of the obtained values for $\Tczero$ was possible due to the analysis of the quasiparticle suppression of $J_s(T)$, which yielded excellent agreement and corroborated the fit of conductivity fluctuations to the resistance data. However, for strongly disordered samples, this sanity check is not possible anymore, since the phase fluctuation regime between $\Tbkt$ and $\Tczero$ becomes comparable to $\Tczero$ itself and thus the temperature evolution of $J_s(T)$ is almost rectangular (Fig. \ref{fig:Js(T)}a). Since $\Tczero$ is a central observable, it is crucial to find an different way of determining $\Tczero$ beyond the three parameter fit of conductivity fluctuations.\\
	Following \cite{Burdastyh2020}, we define $\Tczero$ as inflection point of $R(T)$. The reasoning is the following: Thermally excited free vortices in the regime $\Tbkt<T<\Tczero$ contribute to the resistance according to Eq.\ref{eq:SRC}, which shows strongly positive curvature for typical values of the parameter $b$. Above $\Tczero$, superconducting fluctuations of Aslamazov-Larkin and Maki-Thompson type lead to a negative curvature of $R(T)$. These two regimes meet at the pairing temperature $\Tczero$. Therefore, an inflection point is expected close to $\Tczero$. Fig. \ref{fig:sup:T_c0_analysis} shows an approximately linear segment of $R(T)$, shaded in grey, between two regimes with opposite curvature for two exemplary datasets with different disorder. The first derivative $\mathrm{d}R/\mathrm{d}T$ (green, right $y$-axis) consequently shows a broadened plateau of a few 100 mK. 
	We extract the central point of the linear segment for each film dataset, $T_{center}^{lin}\pm\delta T_{center}^{lin}$ and assume it to be close to the inflection point of the $R(T)$. The error bar $\delta T_{center}^{lin}$ is estimated conservatively to one quarter of the width of the linear segment. $\Tczero$ plotted and discussed in the main text is obtained by defining $\Tczero=T_{center}^{lin}$, where the error bars are smaller than the symbol sizes. An independent fit of the approximate fluctuation conductivity corrections to $R(T)$ for $T\gtrsim \Tczero$ (red line) gives values of $\Tczero$ very close to $T_{center}^{lin}$, indicating robustness of the $\Tczero$-determination.
	
\begin{figure}[h!]
		\centering
		\includegraphics[width=0.7\linewidth]{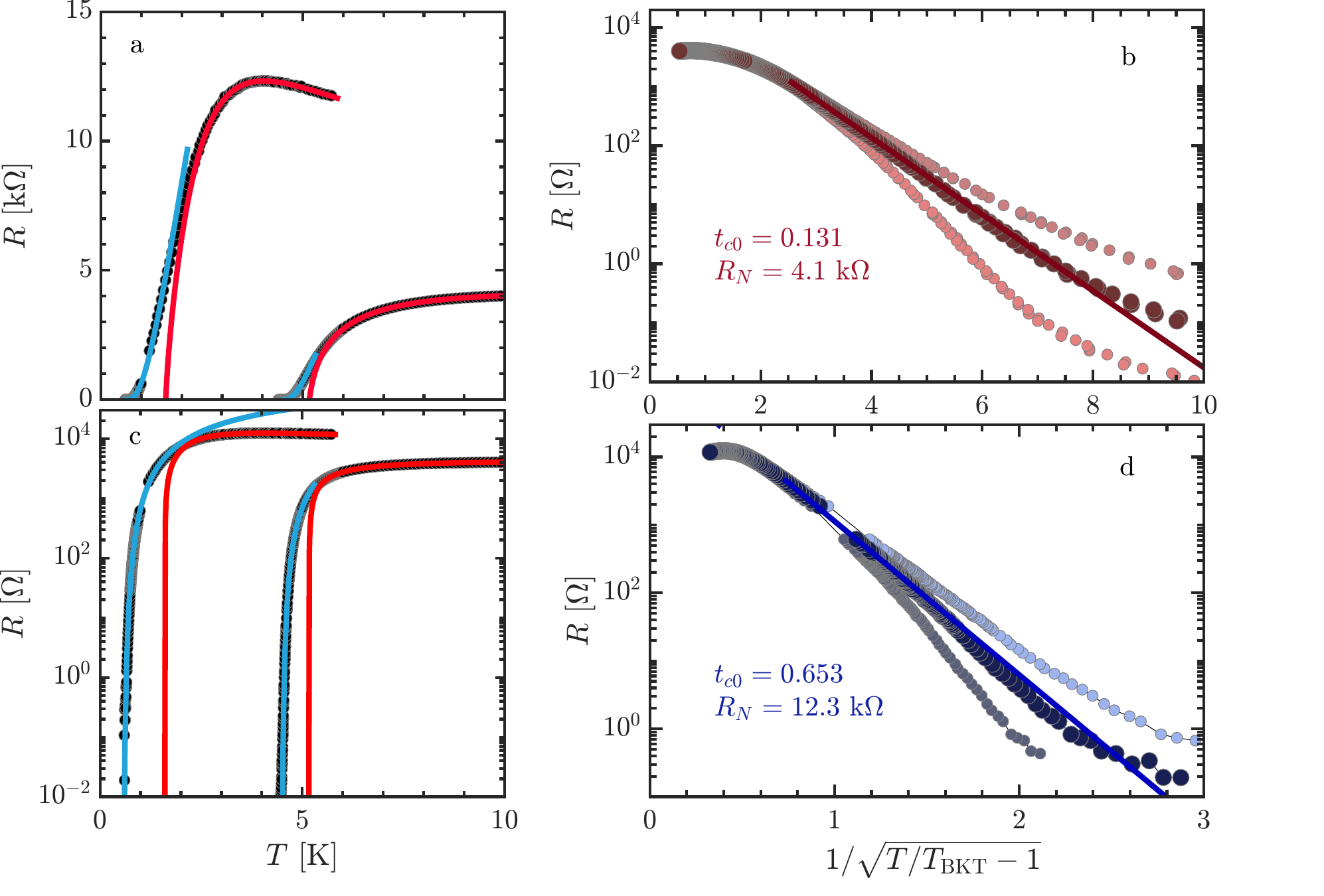}
		\caption[Comparison of transport data with theoretical estimates at moderate and strong disorder]{\textbf{Comparison of transport data with theoretical estimates at moderate and strong disorder} (a), (c): dc transport data of a moderately and strongly disoredered NbN film on logarithmic and linear scale, respectively. Black dots: Experimental data. Red solid lines: Fits to the fluctuation conductivity expression. Blue solid lines: Fits to Eq. \ref{eq:SRC}. (b), (d): data in (a), (c) evaluated separately for the intermediate and high disorder case as a function of reduced temperature $1/\sqrt{T/T_\mathrm{ BKT}-1}$. Large symbols refer to best-fit $T_\mathrm{ BKT}$, small symbols above and below to $T_\mathrm{ BKT}\pm 30$ mK. Solid lines are the fits to Eq. \ref{eq:SRC} shown in (a) and (c). }
		\label{fig:sup:RT_fits_at_high_disorder}
	\end{figure}

	\paragraph{Vortex unbinding temperature $\Tbkt$}:
	
	$\Tbkt$ is assessed from the $R(T)$ datasets by a three-parameter fit of Eq. \ref{eq:SRC} to the data in the regime $T\lesssim \Tczero$, after $\Tczero$ has been obtained according to the procedure laid out in the previous paragraph. The excellent agreement between the values of $\Tbkt$ obtained in this way and the entirely independently determined values from $J_s(T)$ is shown in Fig. \ref{fig:ext_data:bkt_parameters}. Additionally, we show in Fig. \ref{fig:sup:RT_fits_at_high_disorder} the fits on linear and logarithmic scale for two exemplary datasets. The quality of the fit to Eq. \ref{eq:SRC} (shown as blue solid lines in Fig. \ref{fig:sup:RT_fits_at_high_disorder}) can be assessed in detail by plotting experiment and fit as a function of a reduced temperature, $1/\sqrt{T/T_\mathrm{ BKT}-1}$ in Fig. \ref{fig:sup:RT_fits_at_high_disorder}b,d. This display reveals significant deviations from linearity for variations of the fitting parameter $T_\mathrm{ BKT}$ of only $\pm30$ mK for samples both in the intermediate and strong disorder regime, confirming that the central fit-parameter $T_\mathrm{ BKT}$ is constrained to a few percent of $T_\mathrm{ BKT}$.   
	
	\FloatBarrier
	\section{Theoretical Modeling}\label{sec:sup:modeling}
	\subsection{Comparison to analytical models}
	\paragraph{Fermionic suppression of Finkel'stein type}:
	
	\begin{figure}[h!]
		\includegraphics[width=.5\textwidth]{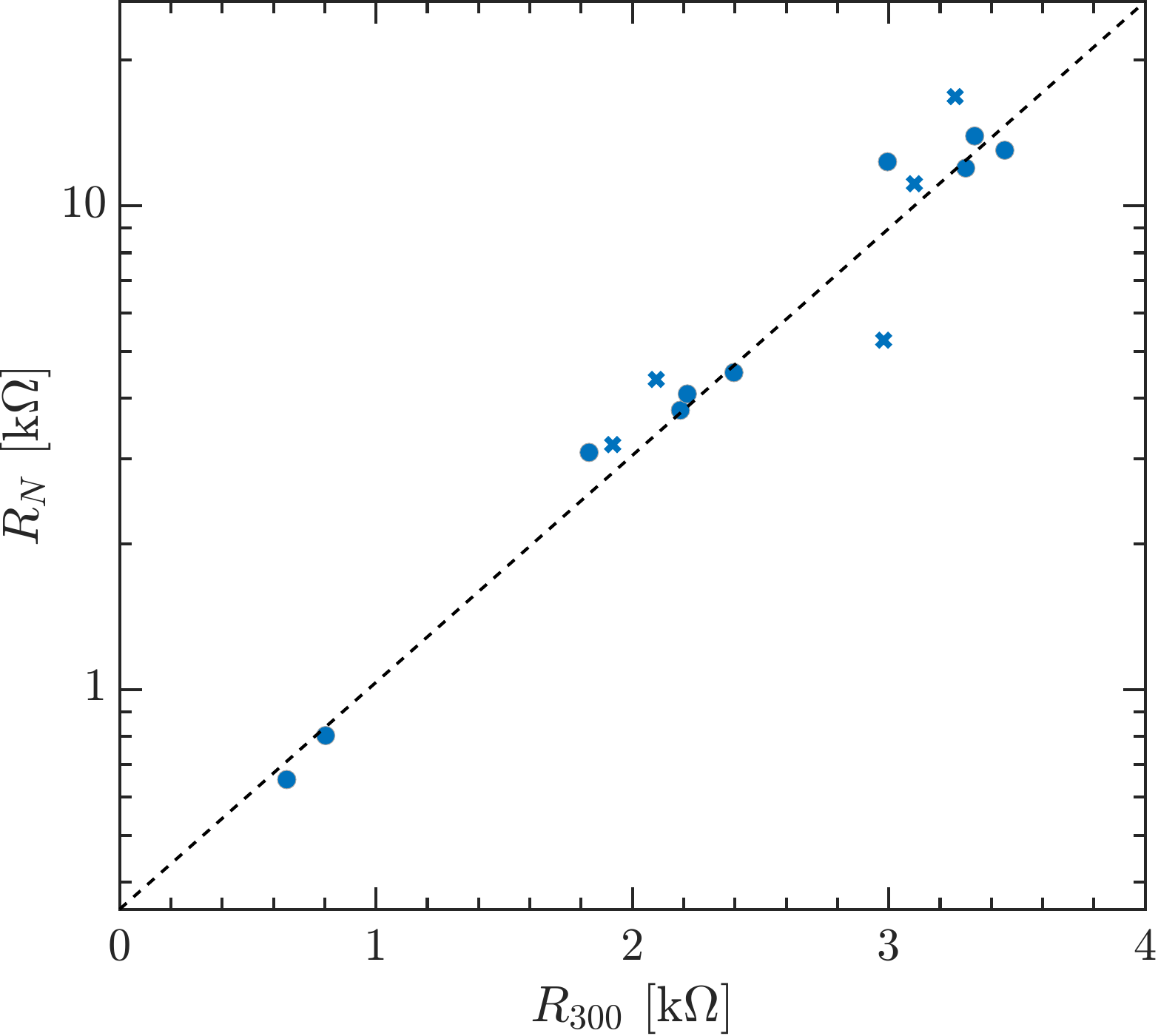}
		\caption[Relationship between room temperature resistance and peak resistance]{Relationship between room temperature resistance and peak resistance. Dotted line is a fit performed on the data indicated by the dots. The crosses represent additional datapoints, which were obtained after the analysis presented in this paper.}
		\label{fig:sup:R_N_vs_R300}
	\end{figure}
	
	The mean-field pairing temperature $\Tczero$ is suppressed due to reduced electronic screening \cite{Finkelshtein1987}:
	
	\begin{align}
		\frac{T_{c0}}{T_\mathrm{c00}}=\exp\left(-\frac{1}{\gamma_c}\right)\left(\frac{\gamma_c+\sqrt{t_D}/2}{\gamma_c-\sqrt{t_D}/2}\right)^{1/\sqrt{t_D}}
		\label{eq:finkelstein}
	\end{align}
	
	where $T_\mathrm{c00}$ is the pairing temperature in the clean case, $\gamma_c=1/\log(T_\mathrm{c00}\tau)$ with mean free time $\tau$ and $t_D=\frac{\pi}{2}\frac{e^2}{h}R_{Drude}$ proportional to the unrenormalized Drude sheet resistance $R_{Drude}$.\\
	Strongly disordered NbN films are subject to quantum corrections to the conductivity (QCC). For decreasing temperature, the resistance increases before dropping to zero at $\Tbkt$. Thus, $R_{300}$, the resistance at $T=300$ K, is taken as a proxy for $R_{Drude}$. The relationship between $R_N$ (the resistance maximum at $T\gtrsim \Tczero$ and $R_{300}$ is roughly exponential and is obtained by a two-parameter least squares fit as shown in Fig. \ref{fig:sup:R_N_vs_R300}.

	The one parameter fit of Eq. \ref{eq:finkelstein} to the $\Tczero(R_{300})$ data is shown in Extended Data Fig. \ref{fig:ext_data:finkelstein}. This fit is also shown as solid light blue line in Fig. \ref{fig:analysis} in the main text, by calculating $R_N(R_{300})$ according to Fig. \ref{fig:sup:R_N_vs_R300}.\\
	
	\subsection{Comparison to Numerics}
	\paragraph{Numerical solution of the Kosterlitz RG equations}:
	
	The fits to the $J_s$-data shown as red curves in Fig. \ref{fig:Js(T)}b,c,e are obtained by numerically solving the Kosterlitz renormalization group equations. The equations read  (\cite{Benfatto2009}):
	
	\begin{align}
		\label{eq:RG_theory}
		\frac{\mathrm{d}K }{\mathrm{d}l}=-K^2g^2,\\
		\frac{\mathrm{d}g}{\mathrm{d}l}=(2-K)g
		\label{eq:RG_theory2}
	\end{align}
	
	with two coupling constants $K$ and $g=2\pi\exp(-\mu/k_BT)$, where $\mu\simeq\frac{\pi^2}{2}J_s$ is the free energy of a vortex-antivortex pair and $l=\log(r/a_0)$ is the dimensionless length scale of a vortex-antivortex pair at separation $r$ and $a_0$ is the ultraviolet cutoff distance of the RG procedure. The superfluid stiffness $J_s$ is obtained as the limit:
	
	\begin{align}
		J_s=\frac{TK(l\rightarrow \infty)}{\pi}
	\end{align}
	
	Numerical solution of the equation system \ref{eq:RG_theory},\ref{eq:RG_theory2}  leads to a flow diagram in the $K-g$ plane. Analyzing the RG flow at different temperatures is equivalent to different initial values for $K(l=0)$ and $g(l=0)$. We choose as starting values the fitted values for the Mattis-Bardeen expression Eq. \ref{eq:J_BCS}, shown as blue (grey) lines in Fig. \ref{fig:Js(T)}b,c (e). This procedure follows Ref. \cite{Benfatto2009}.

	\paragraph{Relationship between $R_N$ and $W$}:
	
	In the main text, Fig. \ref{fig:analysis}, we compare experimental data to numerical modeling based on the attractive Bose-Hubbard model \cite{Stosiek2020}. The Hamiltonian reads: 
	
	Therefore, the tight binding Hamiltonian on a lattice with disorder and attractive interaction is considered:
	

	\begin{align}
		\label{eq:hubbard_hamiltonian}
		\hat{H}_{\text{BdG}} &= \hat{H}_0 + \hat{H}_I \\[6pt]
		\hat{H}_0 &= -t \sum_{\langle i,j \rangle, \sigma} \hat{c}^{\dagger}_{i,\sigma} \hat{c}_{j,\sigma} + \text{H.c.} + \sum_{i=1,\sigma}^{N_{\text{bf}}} (V_i - \mu) \hat{n}_{i,\sigma} \\[6pt]
		\hat{H}_I &= -\frac{U}{2} \sum_{i=1,\sigma}^{N_{\text{bf}}} n(\mathbf{r}_i) \hat{n}_{i,\sigma} - \sum_{i=1}^{N_{\text{bf}}} \Delta(\mathbf{r}_i) \hat{c}_{i,\uparrow} \hat{c}_{i,\downarrow} + \text{H.c.},
	\end{align}
	
	with $t$ the nearest neighbor-hopping constant, $i$, $j$ nearest neighbor sites, $\mu_c$ the chemical potential and the $U$ the on-site pairing interaction. Disorder is introduced by letting ${V}_i$ vary randomly with uniform distribution over $[-{W},{W}]$ for each site. It is important to note that setting $U$ to zero leads to an Anderson insulator and setting $V$ to zero leads to a clean superconductor.
	
\begin{figure}[h]
		\includegraphics[width=.5\textwidth]{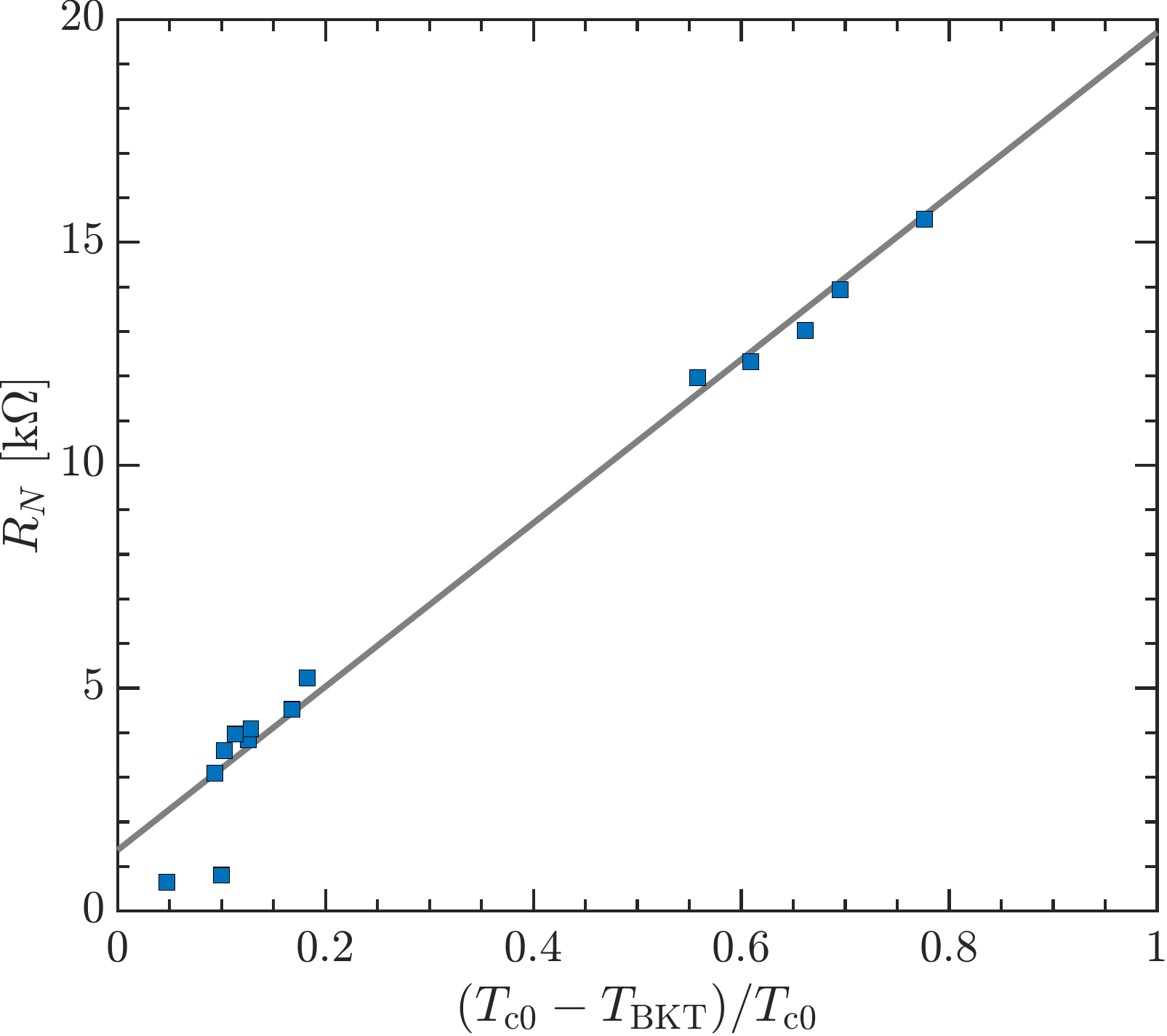}
		\caption[Relationship between peak resistance and reduced temperature]{Relationship between peak resistance and reduced temperature}
		\label{fig:sup:tc0_vs_R_N}
	\end{figure}

	Direct comparison between experiment and numerics is possible by translating $W$, the width of the disorder box distribution, to $R_N$. Experimentally it is found that the distance between mean-field temperature and vortex unbinding temperature depends linearly on $R_N$ (Fig. \ref{fig:sup:tc0_vs_R_N}) in a large range of $R_N$.
	
\begin{figure}[h]
		\includegraphics[width=\textwidth]{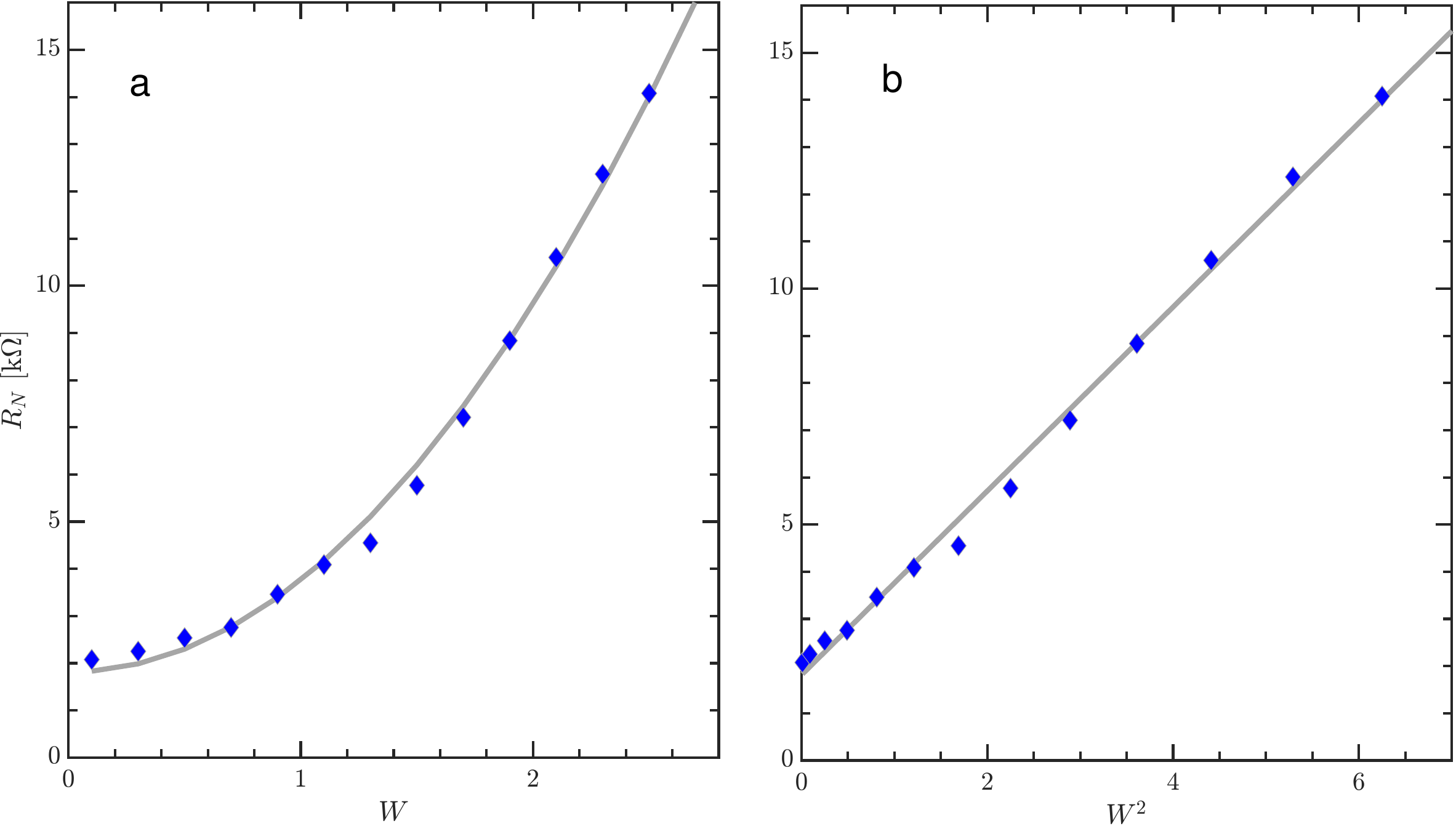}
		\caption[Relationship between Anderson disorder $W$ and peak resistance $R_N$]{Relationship between Anderson disorder $W$ and peak resistance $R_N$. (a) shows $R_N$ vs. $W$, (b) shows $R_N$ vs. $W^2$, illustrating  quadratic behavior. The grey lines represent the same quadratic fit.}
		\label{fig:sup:R_vs_W}
	\end{figure}

	The numerical solution of Eq. \ref{eq:hubbard_hamiltonian} for different disorder strengths directly yields $\Tczero(W)$. From the temperature dependence of the numerically computed $J_s(T, W)$, the vortex-unbinding temperature $\Tbkt(W)$ can be inferred by requiring that $J_s(\Tbkt)$ obeys $J_s(\Tbkt)=2\Tbkt/\pi$ (\cite{Kosterlitz_1974,NelsonKosterlitz_1977}). Thus, the object $(\Tczero-\Tbkt)/\Tczero$ can be extracted as function of disorder from the numerical simulations. Since it is dimensionless, it can be used as a dictionary between numerical and experimental results. Utilizing the linear relationship of $R_N$ and $(\Tczero-\Tbkt)/\Tczero$, experimentally determined $R_N$ and theoretical disorder strength $W$ can be related to each other. Fig. \ref{fig:sup:R_vs_W} demonstrates that $R_N(W)$ scales approximately quadratic, in accordance with expectations in the low-disorder limit \cite{Lee1985-ew}. We believe this procedure approximates $R_N(W)$ well. Caveats in the experimental and numerical determination of the translating object $(\Tczero-\Tbkt)/\Tczero$ arise mainly from three sources. Firstly, the numerical determination of $\Tbkt$ disregards the effects of stiffness renormalization due to vortex-antivortex screening close to $\Tbkt$ and as such systematically overestimates $\Tbkt$ slightly. Secondly, close to the SIT, it appears likely that experimentally determined $J_s$ and $\Tbkt$ for $R_N>10$ k$\Omega$ are affected by quantum phase fluctuations. This effect is not accounted for in the numerical mean field treatment, which reproduces the semi-classical expectation for $J_s$ and thus $\Tbkt$ (see black and dashed grey line Fig. \ref{fig:analysis}. Thirdly, numerically obtained $\Tczero$ does not reflect the experimentally observed suppression of $\Tczero$ due to the insensitivity of the numerical procedure towards Coulomb interactions. The expected inaccuracies  of the first two effects, based on standard BKT RG treatment and the experimentally observed importance of quantum effects on the phase, do not affect the absolute value of $(\Tczero-\Tbkt)/\Tczero$ strongly. The third effect, to a first approximation, applies to $\Tczero$ and $\Tbkt$ equally. A deeper understanding of the relationship between $R_N$ and $W$ remains subject of further study.
	
\begin{thebibliography}{10}
	\expandafter\ifx\csname url\endcsname\relax
	\def\url#1{\texttt{#1}}\fi
	\expandafter\ifx\csname urlprefix\endcsname\relax\def\urlprefix{URL }\fi
	\providecommand{\bibinfo}[2]{#2}
	\providecommand{\eprint}[2][]{\url{#2}}
	
	\bibitem{Sondhi1997-hr}
	\bibinfo{author}{Sondhi, S.~L.}, \bibinfo{author}{Girvin, S.~M.},
	\bibinfo{author}{Carini, J.~P.} \& \bibinfo{author}{Shahar, D.}
	\newblock \bibinfo{title}{Continuous quantum phase transitions}.
	\newblock \emph{\bibinfo{journal}{Rev. Mod. Phys.}}
	\textbf{\bibinfo{volume}{69}}, \bibinfo{pages}{315--333}
	(\bibinfo{year}{1997}).
	
	\bibitem{Gantmakher2010-zs}
	\bibinfo{author}{Gantmakher, V.~F.} \& \bibinfo{author}{Dolgopolov, V.~T.}
	\newblock \bibinfo{title}{Superconductor--insulator quantum phase transition}.
	\newblock \emph{\bibinfo{journal}{Phys.--Usp.}} \textbf{\bibinfo{volume}{53}},
	\bibinfo{pages}{1--49} (\bibinfo{year}{2010}).
	
	\bibitem{Finkelstein1994-ub}
	\bibinfo{author}{Finkel'stein, A.~M.}
	\newblock \bibinfo{title}{Suppression of superconductivity in homogeneously
		disordered systems}.
	\newblock \emph{\bibinfo{journal}{Physica B Condens. Matter}}
	\textbf{\bibinfo{volume}{197}}, \bibinfo{pages}{636--648}
	(\bibinfo{year}{1994}).
	
	\bibitem{Fisher1989-ib}
	\bibinfo{author}{Fisher, M.~P.}, \bibinfo{author}{Weichman, P.~B.},
	\bibinfo{author}{Grinstein, G.} \& \bibinfo{author}{Fisher, D.~S.}
	\newblock \bibinfo{title}{Boson localization and the superfluid-insulator
		transition}.
	\newblock \emph{\bibinfo{journal}{Phys. Rev. B Condens. Matter}}
	\textbf{\bibinfo{volume}{40}}, \bibinfo{pages}{546--570}
	(\bibinfo{year}{1989}).
	
	\bibitem{Fisher1990-cm}
	\bibinfo{author}{Fisher, M.~P.}
	\newblock \bibinfo{title}{Quantum phase transitions in disordered
		two-dimensional superconductors}.
	\newblock \emph{\bibinfo{journal}{Phys. Rev. Lett.}}
	\textbf{\bibinfo{volume}{65}}, \bibinfo{pages}{923--926}
	(\bibinfo{year}{1990}).
	
	\bibitem{Fisher1990-om}
	\bibinfo{author}{Fisher, M.~P.}, \bibinfo{author}{Grinstein, G.} \&
	\bibinfo{author}{Girvin, S.~M.}
	\newblock \bibinfo{title}{Presence of quantum diffusion in two dimensions:
		Universal resistance at the superconductor-insulator transition}.
	\newblock \emph{\bibinfo{journal}{Phys. Rev. Lett.}}
	\textbf{\bibinfo{volume}{64}}, \bibinfo{pages}{587--590}
	(\bibinfo{year}{1990}).
	
	\bibitem{Emery1995-yi}
	\bibinfo{author}{Emery, V.~J.} \& \bibinfo{author}{Kivelson, S.~A.}
	\newblock \bibinfo{title}{Importance of phase fluctuations in superconductors
		with small superfluid density}.
	\newblock \emph{\bibinfo{journal}{Nature}} \textbf{\bibinfo{volume}{374}},
	\bibinfo{pages}{434--437} (\bibinfo{year}{1995}).
	
	\bibitem{Berezinskii1972}
	\bibinfo{author}{Berezinskii, V.~L.}
	\newblock \bibinfo{title}{Destruction of long-range order in one-dimensional
		and two-dimensional systems possessing a continuous symmetry group. ii.
		quantum systems}.
	\newblock \emph{\bibinfo{journal}{Zh. Eksp. Teor. Fiz.}}
	\textbf{\bibinfo{volume}{34}}, \bibinfo{pages}{1144} (\bibinfo{year}{1972}).
	
	\bibitem{Kosterlitz1973-zt}
	\bibinfo{author}{Kosterlitz, J.~M.} \& \bibinfo{author}{Thouless, D.~J.}
	\newblock \bibinfo{title}{Ordering, metastability and phase transitions in
		two-dimensional systems}.
	\newblock \emph{\bibinfo{journal}{J. Phys.}} \textbf{\bibinfo{volume}{6}},
	\bibinfo{pages}{1181--1203} (\bibinfo{year}{1973}).
	
	\bibitem{Kosterlitz1974-mc}
	\bibinfo{author}{Kosterlitz, J.~M.}
	\newblock \bibinfo{title}{The critical properties of the two-dimensional xy
		model}.
	\newblock \emph{\bibinfo{journal}{J. Phys.}} \textbf{\bibinfo{volume}{7}},
	\bibinfo{pages}{1046--1060} (\bibinfo{year}{1974}).
	
	\bibitem{HalperinNelson_1979}
	\bibinfo{author}{Halperin, B.~I.} \& \bibinfo{author}{Nelson, D.~R.}
	\newblock \bibinfo{title}{{Resistive Transition in Superconducting Films}}.
	\newblock \emph{\bibinfo{journal}{{J.~Low Temp.~Phys.}}}
	\textbf{\bibinfo{volume}{36}}, \bibinfo{pages}{599--616}
	(\bibinfo{year}{1979}).
	
	\bibitem{AHNS_1980}
	\bibinfo{author}{V.~Ambegaokar, D. R.~N., B. I.~Halperin} \&
	\bibinfo{author}{Siggia, E.~D.}
	\newblock \bibinfo{title}{Dynamics of superfluid films}.
	\newblock \emph{\bibinfo{journal}{Phys. Rev. B.}}
	\textbf{\bibinfo{volume}{21}}, \bibinfo{pages}{1806--1826}
	(\bibinfo{year}{1980}).
	
	\bibitem{Shahar1992}
	\bibinfo{author}{Shahar, D.} \& \bibinfo{author}{Ovadyahu, Z.}
	\newblock \bibinfo{title}{Superconductivity near the mobility edge}.
	\newblock \emph{\bibinfo{journal}{Phys. Rev. B}} \textbf{\bibinfo{volume}{46}},
	\bibinfo{pages}{10917--10922} (\bibinfo{year}{1992}).
	
	\bibitem{Sacepe2008}
	\bibinfo{author}{Sac\'ep\'e, B.} \emph{et~al.}
	\newblock \bibinfo{title}{Disorder-induced inhomogeneities of the
		superconducting state close to the superconductor-insulator transition}.
	\newblock \emph{\bibinfo{journal}{Phys. Rev. Lett.}}
	\textbf{\bibinfo{volume}{101}}, \bibinfo{pages}{157006}
	(\bibinfo{year}{2008}).
	
	\bibitem{Marrache-Kikuchi2008-hh}
	\bibinfo{author}{Marrache-Kikuchi, C.~A.} \emph{et~al.}
	\newblock \bibinfo{title}{Thickness-tuned superconductor-insulator transitions
		under magnetic field {ina-NbSi}}.
	\newblock \emph{\bibinfo{journal}{Phys. Rev. B Condens. Matter Mater. Phys.}}
	\textbf{\bibinfo{volume}{78}} (\bibinfo{year}{2008}).
	
	\bibitem{Linzen_2017}
	\bibinfo{author}{Linzen, S.} \emph{et~al.}
	\newblock \bibinfo{title}{Structural and electrical properties of ultrathin
		niobium nitride films grown by atomic layer deposition}.
	\newblock \emph{\bibinfo{journal}{Superconductor Science and Technology}}
	\textbf{\bibinfo{volume}{30}}, \bibinfo{pages}{035010}
	(\bibinfo{year}{2017}).
	
	\bibitem{Baturina_2012}
	\bibinfo{author}{Baturina, T.~I.} \emph{et~al.}
	\newblock \bibinfo{title}{Superconducting phase transitions in ultrathin {TiN}
		films}.
	\newblock \emph{\bibinfo{journal}{Europhys.~Lett.}}
	\textbf{\bibinfo{volume}{97}}, \bibinfo{pages}{17012} (\bibinfo{year}{2012}).
	
	\bibitem{Altshuler1982-hs}
	\bibinfo{author}{Altshuler, B.~L.}, \bibinfo{author}{Aronov, A.~G.} \&
	\bibinfo{author}{Khmelnitsky, D.~E.}
	\newblock \bibinfo{title}{Effects of electron-electron collisions with small
		energy transfers on quantum localisation}.
	\newblock \emph{\bibinfo{journal}{J. Phys.}} \textbf{\bibinfo{volume}{15}},
	\bibinfo{pages}{7367--7386} (\bibinfo{year}{1982}).
	
	\bibitem{Koenig2015}
	\bibinfo{author}{K\"onig, E.~J.} \emph{et~al.}
	\newblock \bibinfo{title}{Berezinskii-kosterlitz-thouless transition in
		homogeneously disordered superconducting films}.
	\newblock \emph{\bibinfo{journal}{Phys. Rev. B}} \textbf{\bibinfo{volume}{92}},
	\bibinfo{pages}{214503} (\bibinfo{year}{2015}).
	
	\bibitem{Supplement}
	\bibinfo{title}{Supplementary information}.
	
	\bibitem{Weitzel2023}
	\bibinfo{author}{Weitzel, A.} \emph{et~al.}
	\newblock \bibinfo{title}{Sharpness of the berezinskii-kosterlitz-thouless
		transition in disordered nbn films}.
	\newblock \emph{\bibinfo{journal}{Phys. Rev. Lett.}}
	\textbf{\bibinfo{volume}{131}}, \bibinfo{pages}{186002}
	(\bibinfo{year}{2023}).
	
	\bibitem{Kapitulnik2019-te}
	\bibinfo{author}{Kapitulnik, A.}, \bibinfo{author}{Kivelson, S.~A.} \&
	\bibinfo{author}{Spivak, B.}
	\newblock \bibinfo{title}{Anomalous metals: Failed superconductors}.
	\newblock \emph{\bibinfo{journal}{Rev. Mod. Phys.}}
	\textbf{\bibinfo{volume}{91}}, \bibinfo{pages}{011002}
	(\bibinfo{year}{2019}).
	
	\bibitem{Ghosal1998}
	\bibinfo{author}{Ghosal, A.}, \bibinfo{author}{Randeria, M.} \&
	\bibinfo{author}{Trivedi, N.}
	\newblock \bibinfo{title}{Role of spatial amplitude fluctuations in highly
		disordered s-wave superconductors}.
	\newblock \emph{\bibinfo{journal}{Phys. Rev. Lett.}}
	\textbf{\bibinfo{volume}{81}}, \bibinfo{pages}{3940--3943}
	(\bibinfo{year}{1998}).
	
	\bibitem{Ghosal2001}
	\bibinfo{author}{Ghosal, A.}, \bibinfo{author}{Randeria, M.} \&
	\bibinfo{author}{Trivedi, N.}
	\newblock \bibinfo{title}{Inhomogeneous pairing in highly disordered s-wave
		superconductors}.
	\newblock \emph{\bibinfo{journal}{Phys. Rev. B}} \textbf{\bibinfo{volume}{65}},
	\bibinfo{pages}{014501} (\bibinfo{year}{2001}).
	
	\bibitem{Stosiek2020}
	\bibinfo{author}{Stosiek, M.}, \bibinfo{author}{Lang, B.} \&
	\bibinfo{author}{Evers, F.}
	\newblock \bibinfo{title}{Self-consistent-field ensembles of disordered
		hamiltonians: Efficient solver and application to superconducting films}.
	\newblock \emph{\bibinfo{journal}{Phys. Rev. B}}
	\textbf{\bibinfo{volume}{101}}, \bibinfo{pages}{144503}
	(\bibinfo{year}{2020}).
	
	\bibitem{Charpentier2025}
	\bibinfo{author}{Charpentier, T.} \emph{et~al.}
	\newblock \bibinfo{title}{First-order quantum breakdown of superconductivity in
		an amorphous superconductor}.
	\newblock \emph{\bibinfo{journal}{Nat. Phys.}} \textbf{\bibinfo{volume}{21}},
	\bibinfo{pages}{104--109} (\bibinfo{year}{2025}).
	
	\bibitem{Benfatto2009}
	\bibinfo{author}{Benfatto, L.}, \bibinfo{author}{Castellani, C.} \&
	\bibinfo{author}{Giamarchi, T.}
	\newblock \bibinfo{title}{{Broadening of the Berezinskii-Kosterlitz-Thouless
			superconducting transition by inhomogeneity and finite-size effects}}.
	\newblock \emph{\bibinfo{journal}{Phys. Rev. B}} \textbf{\bibinfo{volume}{80}},
	\bibinfo{pages}{214506} (\bibinfo{year}{2009}).
	
	\bibitem{Maccari2017}
	\bibinfo{author}{Maccari, I.}, \bibinfo{author}{Benfatto, L.} \&
	\bibinfo{author}{Castellani, C.}
	\newblock \bibinfo{title}{{Broadening of the Berezinskii-Kosterlitz-Thouless
			transition by correlated disorder}}.
	\newblock \emph{\bibinfo{journal}{Phys. Rev. B}} \textbf{\bibinfo{volume}{96}},
	\bibinfo{pages}{060508} (\bibinfo{year}{2017}).
	
	\bibitem{Dieplinger}
	\bibinfo{author}{J.~Dieplinger, {\it et al}.}
	\newblock \bibinfo{title}{in preparation}.
	
	\bibitem{WeitzelIII}
	\bibinfo{author}{A.~Weitzel, {\it et al}.}
	\newblock \bibinfo{title}{in preparation} (\bibinfo{year}{2026}).
	
	\bibitem{Larkin2005}
	\bibinfo{author}{Larkin, A.} \& \bibinfo{author}{Varlamov, A.}
	\newblock \emph{\bibinfo{title}{Theory of fluctuations in superconductors}}.
	\newblock International Series of Monographs on Physics
	(\bibinfo{publisher}{Oxford University Press}, \bibinfo{address}{London,
		England}, \bibinfo{year}{2005}).
	
	\bibitem{Caviglia2008-jp}
	\bibinfo{author}{Caviglia, A.~D.} \emph{et~al.}
	\newblock \bibinfo{title}{Electric field control of the {LaAlO3/SrTiO3}
		interface ground state}.
	\newblock \emph{\bibinfo{journal}{Nature}} \textbf{\bibinfo{volume}{456}},
	\bibinfo{pages}{624--627} (\bibinfo{year}{2008}).
	
	\bibitem{Parendo2005-vq}
	\bibinfo{author}{Parendo, K.~A.} \emph{et~al.}
	\newblock \bibinfo{title}{Electrostatic tuning of the superconductor-insulator
		transition in two dimensions}.
	\newblock \emph{\bibinfo{journal}{Phys. Rev. Lett.}}
	\textbf{\bibinfo{volume}{94}}, \bibinfo{pages}{197004}
	(\bibinfo{year}{2005}).
	
	\bibitem{Aubin2006-bg}
	\bibinfo{author}{Aubin, H.} \emph{et~al.}
	\newblock \bibinfo{title}{Magnetic-field-induced quantum
		superconductor-insulator transition {inNb0.15Si0.85}}.
	\newblock \emph{\bibinfo{journal}{Phys. Rev. B Condens. Matter Mater. Phys.}}
	\textbf{\bibinfo{volume}{73}} (\bibinfo{year}{2006}).
	
	\bibitem{Chand2012-kn}
	\bibinfo{author}{Chand, M.} \emph{et~al.}
	\newblock \bibinfo{title}{Phase diagram of the strongly disordereds-wave
		superconductor {NbN} close to the metal-insulator transition}.
	\newblock \emph{\bibinfo{journal}{Phys. Rev. B Condens. Matter Mater. Phys.}}
	\textbf{\bibinfo{volume}{85}} (\bibinfo{year}{2012}).
	
	\bibitem{Mondal_2011a}
	\bibinfo{author}{Mondal, M.} \emph{et~al.}
	\newblock \bibinfo{title}{{Phase Fluctuations in a Strongly Disordered $s$-Wave
			{NbN} Superconductor Close to the Metal-Insulator Transition}}.
	\newblock \emph{\bibinfo{journal}{Phys. Rev. Lett.}}
	\textbf{\bibinfo{volume}{106}}, \bibinfo{pages}{047001}
	(\bibinfo{year}{2011}).
	
	\bibitem{Noat2013}
	\bibinfo{author}{Noat, Y.} \emph{et~al.}
	\newblock \bibinfo{title}{Unconventional superconductivity in ultrathin
		superconducting nbn films studied by scanning tunneling spectroscopy}.
	\newblock \emph{\bibinfo{journal}{Phys. Rev. B}} \textbf{\bibinfo{volume}{88}},
	\bibinfo{pages}{014503} (\bibinfo{year}{2013}).
	
	\bibitem{Carbillet2016}
	\bibinfo{author}{Carbillet, C.} \emph{et~al.}
	\newblock \bibinfo{title}{Confinement of superconducting fluctuations due to
		emergent electronic inhomogeneities}.
	\newblock \emph{\bibinfo{journal}{Phys. Rev. B}} \textbf{\bibinfo{volume}{93}},
	\bibinfo{pages}{144509} (\bibinfo{year}{2016}).
	
	\bibitem{Benfatto_2013}
	\bibinfo{author}{Benfatto, L.}, \bibinfo{author}{Castellani, C.} \&
	\bibinfo{author}{Giamarchi, T.}
	\newblock \bibinfo{title}{{B}erezinskii–{K}osterlitz–{T}houless transition
		within the sine-{G}ordon approach: The role of the vortex-core energy}.
	\newblock \emph{\bibinfo{journal}{40 Years of
			Berezinskii–Kosterlitz–Thouless Theory}} \bibinfo{pages}{161–199}
	(\bibinfo{year}{2013}).
	
	\bibitem{Khvalyuk2024-mr}
	\bibinfo{author}{Khvalyuk, A.~V.}, \bibinfo{author}{Charpentier, T.},
	\bibinfo{author}{Roch, N.}, \bibinfo{author}{Sac{\'e}p{\'e}, B.} \&
	\bibinfo{author}{Feigel'man, M.~V.}
	\newblock \bibinfo{title}{Near power-law temperature dependence of the
		superfluid stiffness in strongly disordered superconductors}.
	\newblock \emph{\bibinfo{journal}{Phys. Rev. B.}}
	\textbf{\bibinfo{volume}{109}} (\bibinfo{year}{2024}).
	
	\bibitem{Levy-Bertrand2019-dy}
	\bibinfo{author}{Levy-Bertrand, F.} \emph{et~al.}
	\newblock \bibinfo{title}{Electrodynamics of granular aluminum from
		superconductor to insulator: Observation of collective superconducting
		modes}.
	\newblock \emph{\bibinfo{journal}{Phys. Rev. B.}} \textbf{\bibinfo{volume}{99}}
	(\bibinfo{year}{2019}).
	
	\bibitem{Sharma2026-eb}
	\bibinfo{author}{Sharma, M.} \emph{et~al.}
	\newblock \bibinfo{title}{Power-law suppression of superfluid stiffness in
		high-kinetic-inductance {NbN} films}  (\bibinfo{year}{2026}).
	\newblock \eprint{2607.20096}.
	
	\bibitem{Baturina2007-du}
	\bibinfo{author}{Baturina, T.~I.}, \bibinfo{author}{Mironov, A.~Y.},
	\bibinfo{author}{Vinokur, V.~M.}, \bibinfo{author}{Baklanov, M.~R.} \&
	\bibinfo{author}{Strunk, C.}
	\newblock \bibinfo{title}{Localized superconductivity in the quantum-critical
		region of the disorder-driven superconductor-insulator transition in {TiN}
		thin films}.
	\newblock \emph{\bibinfo{journal}{Phys. Rev. Lett.}}
	\textbf{\bibinfo{volume}{99}}, \bibinfo{pages}{257003}
	(\bibinfo{year}{2007}).
	
	\bibitem{Sacepe2015-pp}
	\bibinfo{author}{Sac{\'e}p{\'e}, B.} \emph{et~al.}
	\newblock \bibinfo{title}{High-field termination of a cooper-pair insulator}.
	\newblock \emph{\bibinfo{journal}{Phys. Rev. B Condens. Matter Mater. Phys.}}
	\textbf{\bibinfo{volume}{91}} (\bibinfo{year}{2015}).
	
	\bibitem{Poboiko2024-rm}
	\bibinfo{author}{Poboiko, I.} \& \bibinfo{author}{Feigel'man, M.~V.}
	\newblock \bibinfo{title}{Mean-field theory of first-order quantum
		superconductor-insulator transition}.
	\newblock \emph{\bibinfo{journal}{SciPost Phys.}} \textbf{\bibinfo{volume}{17}}
	(\bibinfo{year}{2024}).
	
	\bibitem{Mondal_2011b}
	\bibinfo{author}{Mondal, M.} \emph{et~al.}
	\newblock \bibinfo{title}{{Role of the Vortex-Core Energy on the
			Berezinskii-Kosterlitz-Thouless Transition in Thin Films of NbN}}.
	\newblock \emph{\bibinfo{journal}{Phys. Rev. Lett.}}
	\textbf{\bibinfo{volume}{107}}, \bibinfo{pages}{217003}
	(\bibinfo{year}{2011}).
	
	\bibitem{Baumgartner_2020}
	\bibinfo{author}{Baumgartner, C.} \emph{et~al.}
	\newblock \bibinfo{title}{{Josephson Inductance as a Probe for Highly Ballistic
			Semiconductor-Superconductor Weak Links}}.
	\newblock \emph{\bibinfo{journal}{Phys. Rev. Lett.}}
	\textbf{\bibinfo{volume}{126}}, \bibinfo{pages}{037001}
	(\bibinfo{year}{2021}).
	
	\bibitem{Yong2013}
	\bibinfo{author}{Yong, J.}, \bibinfo{author}{Lemberger, T.~R.},
	\bibinfo{author}{Benfatto, L.}, \bibinfo{author}{Ilin, K.} \&
	\bibinfo{author}{Siegel, M.}
	\newblock \bibinfo{title}{{Robustness of the Berezinskii-Kosterlitz-Thouless
			transition in ultrathin NbN films near the superconductor-insulator
			transition}}.
	\newblock \emph{\bibinfo{journal}{Phys. Rev. B}} \textbf{\bibinfo{volume}{87}},
	\bibinfo{pages}{184505} (\bibinfo{year}{2013}).
	
	\bibitem{Mandal2020}
	\bibinfo{author}{Mandal, S.} \emph{et~al.}
	\newblock \bibinfo{title}{{Destruction of superconductivity through phase
			fluctuations in ultrathin $a$-MoGe films}}.
	\newblock \emph{\bibinfo{journal}{Phys. Rev. B}}
	\textbf{\bibinfo{volume}{102}}, \bibinfo{pages}{060501}
	(\bibinfo{year}{2020}).
	
	\bibitem{Hebard1983-cw}
	\bibinfo{author}{Hebard, A.~F.} \& \bibinfo{author}{Fiory, A.~T.}
	\newblock \bibinfo{title}{Critical-exponent measurements of a two-dimensional
		superconductor}.
	\newblock \emph{\bibinfo{journal}{Phys. Rev. Lett.}}
	\textbf{\bibinfo{volume}{50}}, \bibinfo{pages}{1603--1606}
	(\bibinfo{year}{1983}).
	
	\bibitem{Pappas2011-hx}
	\bibinfo{author}{Pappas, D.~P.}, \bibinfo{author}{Vissers, M.~R.},
	\bibinfo{author}{Wisbey, D.~S.}, \bibinfo{author}{Kline, J.~S.} \&
	\bibinfo{author}{Gao, J.}
	\newblock \bibinfo{title}{Two level system loss in superconducting microwave
		resonators}.
	\newblock \emph{\bibinfo{journal}{IEEE Trans. Appl. Supercond.}}
	\textbf{\bibinfo{volume}{21}}, \bibinfo{pages}{871--874}
	(\bibinfo{year}{2011}).
	
	\bibitem{LorenzDiss}
	\bibinfo{author}{Fuchs, L.}
	\newblock \emph{\bibinfo{title}{Interplay of Spin-Orbit Interaction and
			Two-Dimensional Superconductivity in Al/InAs Heterostructures}}.
	\newblock \bibinfo{type}{Phd thesis}, \bibinfo{school}{Universit{\"a}t
		Regensburg}, \bibinfo{address}{Universität Regensburg}
	(\bibinfo{year}{2021}).
	\newblock \bibinfo{note}{Available at
		\url{https://epub.uni-regensburg.de/50819}}.
	
	\bibitem{Glatz2011-af}
	\bibinfo{author}{Glatz, A.}, \bibinfo{author}{Varlamov, A.~A.} \&
	\bibinfo{author}{Vinokur, V.~M.}
	\newblock \bibinfo{title}{Fluctuation spectroscopy of disordered
		two-dimensional superconductors}.
	\newblock \emph{\bibinfo{journal}{Phys. Rev. B Condens. Matter Mater. Phys.}}
	\textbf{\bibinfo{volume}{84}} (\bibinfo{year}{2011}).
	
	\bibitem{Burdastyh2020}
	\bibinfo{author}{Burdastyh, M.~V.} \emph{et~al.}
	\newblock \bibinfo{title}{Superconducting phase transitions in disordered
		{NbTiN} films}.
	\newblock \emph{\bibinfo{journal}{Sci. Rep.}} \textbf{\bibinfo{volume}{10}},
	\bibinfo{pages}{1471} (\bibinfo{year}{2020}).
	
	\bibitem{Finkelshtein1987}
	\bibinfo{author}{{Finkel'stein, A M}, A.~M.}
	\newblock \bibinfo{title}{{Superconductivity-transition temperature in
			amorphous films}}.
	\newblock \emph{\bibinfo{journal}{Pisma v Zhurnal Eksperimentalnoi i
			Teoreticheskoi Fiziki}} \textbf{\bibinfo{volume}{45}},
	\bibinfo{pages}{37--40} (\bibinfo{year}{1987}).
	
	\bibitem{Kosterlitz_1974}
	\bibinfo{author}{Kosterlitz, J.~M.}
	\newblock \bibinfo{title}{The critical properties of the two-dimensional
		xy-model}.
	\newblock \emph{\bibinfo{journal}{J. Phys. C: Solid State Phys.}}
	\textbf{\bibinfo{volume}{7}}, \bibinfo{pages}{1046} (\bibinfo{year}{1974}).
	
	\bibitem{NelsonKosterlitz_1977}
	\bibinfo{author}{Nelson, D.~R.} \& \bibinfo{author}{Kosterlitz, J.~M.}
	\newblock \bibinfo{title}{{Universal Jump in the Superfluid Density of
			Two-Dimensional Superfluids}}.
	\newblock \emph{\bibinfo{journal}{Phys. Rev. Lett.}}
	\textbf{\bibinfo{volume}{39}}, \bibinfo{pages}{1201--1205}
	(\bibinfo{year}{1977}).
	
	\bibitem{Lee1985-ew}
	\bibinfo{author}{Lee, P.~A.} \& \bibinfo{author}{Ramakrishnan, T.~V.}
	\newblock \bibinfo{title}{Disordered electronic systems}.
	\newblock \emph{\bibinfo{journal}{Rev. Mod. Phys.}}
	\textbf{\bibinfo{volume}{57}}, \bibinfo{pages}{287--337}
	(\bibinfo{year}{1985}).
	
\end{thebibliography}
\end{document}